\documentclass[journal]{IEEEtran}
\usepackage{amssymb}
\usepackage[cmex10]{amsmath}
\usepackage{stfloats}
\usepackage{graphicx}
\usepackage{subfigure}
\usepackage{tabularx}
\usepackage{epsfig,epsf,color,balance,cite}
\usepackage{verbatim}
\usepackage{url}
\usepackage{bm}

\usepackage{lipsum}

\newtheorem{theorem}{Theorem}
\newtheorem{coro}{Corollary}

\usepackage{algorithm}
\usepackage{algpseudocode}

\begin{document}

 \title{Power Control for Multi-Cell Networks with Non-Orthogonal Multiple Access}
\author{
\IEEEauthorblockN{Zhaohui Yang, 
Cunhua Pan,  \IEEEmembership{Member, IEEE},
Wei Xu,  \IEEEmembership{Senior Member, IEEE},
Yijin Pan,  
Ming Chen,  \IEEEmembership{Member, IEEE},
and Maged Elkashlan,  \IEEEmembership{Member, IEEE}
}

\thanks{Z. Yang, W. Xu, Y. Pan and M. Chen are with the National Mobile Communications Research Laboratory, Southeast University, Nanjing 211111, China, Email: \{yangzhaohui, wxu, panyijin, chenming\}@seu.edu.cn.}
\thanks{C. Pan and M. Elkashlan are with School of Electronic Engineering and Computer Science, Queen Mary University of London, London E1 4NS, U.K.,
 Email: \{c.pan, maged.elkashlan\}@qmul.ac.uk.
}}

\maketitle

\begin{abstract}
In this paper, we investigate the problems of sum power minimization and sum rate maximization for multi-cell networks with non-orthogonal multiple access. Considering the sum power minimization, we obtain closed-form solutions to the optimal power allocation strategy and then successfully transform the original problem to a linear one with a much smaller size, which can be optimally solved by using the standard interference function. To solve the nonconvex sum rate maximization problem, we first prove that the power allocation problem for a single cell is a convex problem. By analyzing the Karush-Kuhn-Tucker conditions, the optimal power allocation for users in a single cell is derived in closed form. Based on the optimal solution in each cell, a distributed algorithm is accordingly proposed to acquire efficient solutions. Numerical results verify our theoretical findings showing the superiority of our solutions compared to the orthogonal frequency division multiple access and broadcast channel.
\end{abstract}

\IEEEpeerreviewmaketitle


\section{Introduction}
With the explosive growth of data traffic in mobile Internet \cite{thompson20145g}, non-orthogonal multiple access (NOMA) has been recently proposed \cite{7263349,Zhiguo2017Survey}.
By using superposition coding at the transmitter and successive interference
cancelation (SIC) at the receiver, NOMA can achieve higher spectral efficiency than conventional orthogonal multiple access, such as orthogonal frequency division multiple access (OFDMA).
Besides, NOMA can support more connections by letting more than one user simultaneously access the same frequency or time resources \cite{6933472,7462187,7551125,7731144}.
Therefore, NOMA has been deemed as a promising multiple access scheme for the next generation mobile communication networks \cite{saito2013non,ding2017application,7236924}.

NOMA can simultaneously serve multiple users with the same frequency band and time slot by splitting them in the power domain \cite{7676258}.
The basic concept of NOMA with SIC receiver was introduced in \cite{6666209}.
The ergodic sum rate and the outage performance with fixed power allocation  were analyzed in \cite{ding2014performance} for NOMA.
Specific impacts of power allocation on sum rate \cite{6824963,7277111,7890454}, fairness \cite{timotheou2015fairness,7022998,Yang2016Fair}, and energy efficiency \cite{zhang2016noma,Yang2017Energy} were investigated.
Moreover, joint sub-channel and power allocation was studied in \cite{7587811,7063568,7523951,7560605,7562275}.
However, the above existing works \cite{6666209,ding2014performance,6824963,7277111,7890454,timotheou2015fairness,7022998,Yang2016Fair,zhang2016noma,7587811,7063568,7523951,7560605,7562275,Yang2017Energy} are limited to single-cell analysis, where there is no inter-cell interference.


Recently, NOMA has been extended to multi-cell networks \cite{6708131,7582424,shin2016non,han2014energy,7504068,7510796,Nguyen2017PrecoderNOMA}.
Different from multi-cell networks with OFDMA, users in the same cell can receive intra-cell interference, which can be subtracted by using SIC in multi-cell networks with NOMA.
Different from single-cell networks with NOMA, inter-cell interference should be considered in multi-cell networks with NOMA.
To harness the effect of inter-cell interference, joint processing (JP) and coordinated scheduling (CS) technologies are usually adopted \cite{6146494}.
In NOMA-JP, the users' data symbols are available at more than one base station (BS) \cite{6708131}.
The designs of CS for NOMA differ from those of JP in that the users' data are not shared among the BSs \cite{7582424}.
However, the cooperating BSs in NOMA-CS still need to exchange global channel state information and cooperative scheduling information via a standardized interface named X2.
In addition, the number of transmission points is one for NOMA-CS \cite{7582424}, while  the number of transmission points is more than one for NOMA-JP in \cite{6708131}.
Moreover, the numbers of supported users by NOMA-CS and NOMA-JP are different \cite{shin2016non}.

Based on NOMA-CS, the uplink and downlink power control problems for sum power minimization in two-cell networks with NOMA were studied in \cite{7504068} and \cite{7510796}, respectively.
However,
the above existing works \cite{han2014energy,6708131,7510796,7504068} were merely limited to power minimization in a two-cell network with only two users per cell.
The sum rate maximization problem in multiple-input multiple-output (MIMO)-NOMA multi-cell networks was investigated in \cite{Nguyen2017PrecoderNOMA}, where two users are paired in a virtual cluster.
Consequently,
there is a lack of systematic approach for sum power minimization and sum rate maximization problems via power control in multi-cell networks with NOMA from a mathematical optimization point of view.

The objective of this paper is to investigate the power control problems for sum power minimization and sum rate maximization in multi-cell networks with NOMA.
The contributions of this paper are summarized as follows:

\begin{enumerate}
  \item  The sum power minimization problem can be equivalently transformed into a linear problem (LP) with smaller variables.
Having obtained the total transmission power of all BSs, the power allocation for each user can be presented in a closed-form expression.
  \item  To solve the sum rate maximization problem, we decouple it into two subproblems, i.e., power allocation problem for users in a single cell and power control problem in multiple cells.
 With the total transmission power of all BSs fixed, we succeed in showing that power control problem in a single cell is a convex problem.
 By analyzing the Karush-Kuhn-Tucker (KKT) conditions, we observe that
only the user with the highest channel gain deserves additional power allocation, while the power for other users in the same cell is merely determined to maintain their minimal rate demands.
Closed-form expressions of the optimal power allocation and optimal sum rate in a single cell are further obtained.
\item Based on the closed-form expression of the optimal sum rate in a single cell, the original sum rate maximization problem can be equivalently transformed into a simplified problem.
    Since the objective function for each BS can be formulated as a difference of two convex functions (DC),
    a convex approximation of the objective function is introduced.
    Furthermore, a distributed algorithm is also provided to obtain a suboptimal solution.
\end{enumerate}

The rest of the paper is organized as follows.
In Section $\text{\uppercase\expandafter{\romannumeral2}}$, we introduce the system model and power control formulation.
Sum power minimization and sum rate maximization for multi-cell networks with NOMA are addressed in Section $\text{\uppercase\expandafter{\romannumeral 3}}$ and Section $\text{\uppercase\expandafter{\romannumeral 4}}$, respectively.
The extension to MIMO-NOMA systems is introduced in Section $\text{\uppercase\expandafter{\romannumeral5}}$.
Some numerical results are shown in Section $\text{\uppercase\expandafter{\romannumeral6}}$
and conclusions are finally drawn in Section $\text{\uppercase\expandafter{\romannumeral7}}$.

\section{System Model and Problem Formulation}

Consider a downlink multi-cell network with NOMA, 
where there are $I$ BSs and $J$ users. Denote the set of BSs and users by ${\mathcal{I}}=\{1, 2, \cdots, I\}$ and ${\mathcal{J}}=\{1, 2, \cdots, J\}$, respectively.
The unique group of users served by BS $i\in {\mathcal{I}}$ is denoted by set ${\cal{J}}_i=\{J_{i-1}+1,J_{i-1}+2,\cdots, J_{i}\}$, where $J_0=0$, $J_I=J$, $J_i=\sum_{l=1}^i|{\cal{J}}_l|$, and $|\cdot|$ is the cardinality of a set.
We focus on the downlink where mutual interference exists among cells.

By using NOMA, a BS serves multiple users by splitting them in the power domain.
Assume that each BS shares the same spectrum.
For each BS, the total bandwidth, $B_{\text{total}}$ is equally divided into $M$ subchannels, where the bandwidth of each subchannel is $B=B_{\text{total}}/M$.
Let $\mathcal M=\{1, 2, \cdots, M\}$ be the set of subchannels.
The set of users served by BS $i$ on subchannel $m$ is denoted by $\mathcal J_{im}=\{J_{i(m-1)}+1, J_{i(m-1)}+2, \cdots, J_{im}\}$, where $J_{i0}=J_{i-1}+1$, $J_{iM}=J_i$, $J_{im}=\sum_{l=1}^m |\mathcal J_{il}|$ and $|{\cal{J}}_{il}|\geq 2$.
The channel gain between BS $i$ and user $j\in\mathcal J_{im}$ on subchannel $m$ is denoted by $h_{ijm}$.
Without loss of generality, the channels are sorted as $|h_{i(J_{i(m-1)}+1)m}|\leq |h_{i(J_{i(m-1)}+2)m}| \leq\cdots\leq |h_{iJ_{im}m}|$, $\forall i \in \mathcal I$, $m\in\mathcal M$.
According to the NOMA principle,  BS $i$ simultaneously transmits signal $s_{im}$ to all its served users in $\mathcal J_{im}$.
The transmitted signal $s_{im}$ can be expressed as
\begin{equation}\label{sys1}
s_{im}=\sum_{j=J_{i(m-1)}+1}^{J_{im}} \sqrt{p_{ijm}}s_{ijm},
\end{equation}
where $s_{ijm}$ and $p_{ijm}$ are the message and allocated power for user $j\in\mathcal J_{im}$, respectively.

The observation at user $j\in \mathcal J_{im}$ on subchannel $m$ is given by
\begin{eqnarray}\label{sys2}
y_{ijm}
&&\!\!\!\!\!\!\!\!\!\!\!
=h_{ijm} s_{im}+\sum_{k \in \mathcal I \setminus \{i\}} h_{kjm}  s_{km}+n_{jm}
\nonumber\\
&&\!\!\!\!\!\!\!\!\!\!\!
=\sum_{l=J_{i(m-1)}+1}^{J_{im}}h_{ijm} \sqrt{p_{ilm}}  s_{ilm} \nonumber\\
&&\!\!\!\!\!\!\!\!\!\!\!  +
\sum_{k \in \mathcal I \setminus \{i\}}\sum_{n=J_{k(m-1)}+1}^{J_{km}}h_{kjm} \sqrt{p_{knm}}  s_{knm} +n_{jm},
\end{eqnarray}
where $h_{kjm}$ is the cross channel gain between BS $k$ and user $j$ served by BS $i$ on subchannel $m$, and $n_{jm}$ represents the additive zero-mean Gaussian noise with variance $\sigma^2$.
According to \cite{6666209,ding2014performance,6824963,7277111}, each user should decode the messages of other users in the same cell with lower channel gains before decoding its own message.
Denoting the total transmission power of BS $i$ on subchannel $m$ by $q_{im}=\sum_{j=J_{i(m-1)}+1}^{J_{im}}p_{ijm}$,
the achievable rate of user $j\in\mathcal J_{im}$ to detect the message of user $l\in\{J_{i(m-1)}+1, \cdots, j\}$ on subchannel $m$ is
\begin{equation}\label{sys2_2}
r_{ijlm} = B\log_2\left(
1+\frac{|h_{ijm}|^2 p_{ilm}}
{|h_{ijm}|^2 \sum_{n=l+1}^{J_{im}} p_{inm}+Z_{ijlm}}
\right),
\end{equation}
where
\begin{equation}
Z_{ijlm}=
\sum_{k \in \mathcal I \setminus \{i\}} q_{km} |h_{kjm}|^2+\sigma^2.
\end{equation}

According to (\ref{sys2_2}), strong user $j$ with high channel gain needs to decode the message of weak user $l\leq j$ with low channel gain.
To ensure successful SIC,
the achievable rate of user $j\in\mathcal J_{im}$ on subchannel $m$ can be given by\footnote{As in \cite{7277111}, the decoding order is determined by the increasing order of channel gains for users in the same cell. The optimal decoding order for multi-cell networks is still an open problem\cite{shin2016non}, which is beyond the scope of this paper.}
\begin{eqnarray}
r_{ijm}&&\!\!\!\!\!\!\!\!\!=\min_{l\in\{j, \cdots, J_{im}\}} r_{iljm}\nonumber\\
&&\!\!\!\!\!\!\!\!\!=\min_{l\in\{j, \cdots, J_{im}\}}B\log_2\left(
1+\frac{p_{ijm}}
{ \sum_{n=j+1}^{J_{im}} p_{inm}+
\frac{Z_{iljm}}{|h_{ilm}|^2}}
\right)
\nonumber\\
&&\!\!\!\!\!\!\!\!\!=
B\log_2\left(
1+\frac{p_{ijm}}
{ \sum_{n=j+1}^{J_{im}} p_{inm}+
H_{ijm}}
\right),\label{sys3}
\end{eqnarray}
where
\begin{equation}\label{sys3_2}
H_{ijm}=\max_{l\in\{j, \cdots, J_{im}\}}\frac{\sum_{k \in \mathcal I \setminus \{i\}} q_{km} |h_{klm}|^2+\sigma^2}{|h_{ilm}|^2}.
\end{equation}

Denote $R_{ijm}>0$ as the minimal rate demand of user $j\in \mathcal J_{im}$ on subchannel $m$.
Applying (\ref{sys3}), $r_{ijm} \geq R_{ijm}$ is equivalent to the following linear constraint:
\begin{equation}\label{sys5}
 p_{ijm} \geq
\left(2^{\frac{R_{ijm}} B }-1\!\right) \left(
 \sum_{n=j+1}^{J_{im}} p_{inm}+
H_{ijm}
 \right).
\end{equation}

Our objective is to optimize the power allocation in order to minimize the sum power or maximize the sum rate under the total power constraints and individual rate demands.
Mathematically, the power control problem can be formulated as
\begin{subequations}\label{max1}
\begin{align}
\mathop{\min}_{\pmb{p}\geq \pmb 0, \pmb q\geq\pmb 0}\;
 \quad&   V(\pmb p, \pmb q)\\
\textrm{s.t.}\quad\qquad \!\!\!\!\!\!\!\!\!
&
q_{im}=\sum_{j=J_{i(m-1)}+1}^{J_{im}}p_{ijm}, \quad \forall i \in \mathcal I, m \in \mathcal M \\
&
p_{ijm} \geq
\left(2^{\frac{R_{ijm}} B }-1\!\right) \left(
 \sum_{n=j+1}^{J_{im}} p_{in}+
H_{ijm}
 \right),
 \nonumber \\
 &\qquad
 \forall i \in \mathcal I, m \in\mathcal M, j\in \mathcal J_{im}
 \\
 &\sum_{m=1}^M q_{im}\leq Q_i, \quad\forall i \in \mathcal I,
\end{align}
\end{subequations}
where $\pmb p =[p_{111}, \cdots, p_{1J_{1M}{M}}, \cdots, p_{IJ_{IM}M}]^T$ is the transmission power vector, $\pmb q=[q_{11}, \cdots,$ $q_{1M}, \cdots, q_{IM}]^T$ is the total transmission power vector,
$H_{ijm}$ is defined in (\ref{sys3_2}),
and $Q_{i}$ is the maximum transmission power of the BS $i$.
Constraints (\ref{max1}c) reflect that the minimal rate demands of all users can be satisfied.
$V(\pmb p, \pmb q)$ is the objective function, which can be sum power $\sum_{i=1}^I \sum_{m=1}^M q_{im}$ or negative sum rate $-\sum_{i =1}^I\sum_{m=1}^M \sum_{j=J_{i(m-1)}+1}^{J_{im}} r_{ijm}$ with $r_{ijm}$ defined in (\ref{sys3}).

Obviously, the feasible set of Problem (\ref{max1}) is linear.
For sum power minimization, Problem (\ref{max1}) is a LP, of which the globally optimal solution can be effectively obtained.
In the following, we show that sum power minimization problem can  be equivalently transformed into a smaller LP.
Since the problem of sum rate maximization problem is nonconvex, obtaining global optimum however is known to be difficult.
To solve the sum rate maximization problem efficiently, we first consider the power allocation problem for users in a single cell with fixed total transmission power of all BSs.
Then, based on the optimal power allocation for users in a single cell, the primal multi-cell sum rate maximization problem can also be simplified into an equivalent problem.
A distributed algorithm is proposed to obtain a suboptimal solution of the simplified sum rate maximization problem.
\section{Sum Power Minimization for Multi-Cell Networks}
In this section,  we solve the sum power minimization Problem (\ref{max1}) with
\begin{equation}\label{power_mini1_obj1}
V(\pmb p, \pmb q)=\sum_{i=1}^{I}\sum_{m=1}^M q_{im}.
\end{equation}
Obviously, sum power minimization Problem (\ref{max1}) is a LP.
According to (\ref{sys5}), the interference level received by each user is determined by total transmission power $\pmb q$.
Once the total transmission power of other BSs is given, Problem (\ref{max1}) with objective function (\ref{power_mini1_obj1}) can be simplified to a single-cell power minimization problem with fixed inter-cell interference, which fortunately has closed-form solution.
Substituting the closed-form solution for each cell into Problem (\ref{max1}), we can obtain an equivalent LP with a much smaller size, i.e., far fewer variables.
\subsection{Distributed Power Control Algorithm}
\begin{theorem}\label{power_mini1}
For sum power minimization Problem (\ref{max1}) with objective function (\ref{power_mini1_obj1}), power $\pmb p$ can be optimally solved with closed-form expression as
\begin{eqnarray}\label{power_mini2}
p_{ijm}&&\!\!\!\!\!\!\!\!\!\!\!  =
\sum_{n=j}^{J_{im}}
{ \left(2^{\frac{R_{inm}}B} -1\right)
{2^{\sum_{s=j}^{n-1}\frac{R_{ism}}B}}H_{inm}}
\nonumber  \\
&&\!\!\!\!\!\!\!\!\!\!\! \quad-
\sum_{n=j+1}^{J_{im}}
{ \left(2^{\frac{R_{inm}}B} \!-\!1\!\right)
{2^{\sum_{s=j+1}^{n-1}\frac{R_{ism}}B}}H_{inm}}, 
\end{eqnarray}
while the optimal solution to $\pmb q$ (for determining $H_{inm}$) is determined by solving a smaller LP in (\ref{max2}) with much lower complexity.
\begin{subequations}\label{max2}
\begin{align}
\mathop{\min}_{\pmb q\geq \pmb 0}\;
 \quad&  \sum_{i=1}^I \sum_{m=1}^M q_{im}\\
\textrm{s.t.}\quad\qquad \!\!\!\!\!\!\!\!\!
&
q_{im}\geq \sum_{j=J_{i(m-1)}+1}^{J_{im}}\left(2^{\frac{R_{ijm}}B} -1\right)
{2^{\sum_{s=J_{i(m-1)}+1}^{j-1}\frac{R_{ism}}B}}
\nonumber\\
&\qquad\!\!\!\!\!\!\!
\max_{l\in\{j, \cdots, J_{im}\}}\frac{\sum_{k \in \mathcal I \setminus \{i\}} q_{km} |h_{klm}|^2+\sigma^2}{|h_{ilm}|^2}
\nonumber\\
&\qquad\!\!\!\!\!\!\!\:\;\triangleq f_{im}(\pmb q),
\quad\forall i  \in  \mathcal I, m \in \mathcal M\\
&\sum_{m=1}^M q_{im}\leq Q_i, \quad\forall i \in \mathcal I.
\end{align}
\end{subequations}
\end{theorem}

\itshape \textbf{Proof:}  \upshape
Please refer to Appendix A.
 \hfill $\Box$

Note that the concept of strong user reflected from (\ref{sys2_2}), (\ref{sys3}) and (\ref{sys3_2}) is helpful in obtaining the closed-form expression in (\ref{power_mini2}).
It can be verified that the optimal solution to LP (\ref{max2}) with given $[q_{1m}, \cdots, q_{(i-1)m}, q_{(i+1)m}, \cdots, q_{Im}]^T$ can be directly obtained by $q_{im}=f_{im}(\pmb q)$.
To solve LP (\ref{max2}) for multi-cell power minimization, we provide a distributed algorithm in Algorithm 1.

\begin{algorithm}[h]
\caption{Distributed Power Control for Sum Power Minimization (DPC-SPM)}
\label{alg:Framwork1}
\begin{algorithmic}[1]
\State Initialize $q_{im}^{(0)}=Q_i/M$, $\forall i \in \mathcal N, m \in\mathcal M$.
Set $t=1$, and maximal iteration number $T_{\max}$.
\For{$i=1, 2, \cdots, I$}
\For{$m=1, 2, \cdots, M$}
\State Calculate $q_{im}^{(t)}=f_{im}(\pmb q^{(t-1)})$;
\EndFor
\EndFor
\State
If $t > T_{\max}$ or objective function (\ref{max2}a) converges,
output $\pmb q^*=\pmb q^{(t)}$, and
terminate.
Otherwise,
$\pmb q^{(t)}=[q_{11}^{(t)}, \cdots, q_{1M}^{(t)}, \cdots, q_{IM}^{(t)}]^T$,
set $t=t+1$ and go to step 2.
\end{algorithmic}
\end{algorithm}
\subsection{Convergence and Global Optimality}
To show the convergence and global optimality of DPC-SPM algorithm,
we recap the standard interference function introduced in \cite{yates1995framework}.\footnote{
Consider an arbitrary interference function $\pmb{D}(\pmb{q})=[D_{11}(\pmb{q}), \cdots, D_{IM}(\pmb{q})]^T$, we say $\pmb{D}(\pmb{q})$ is a standard interference function if for all $\pmb{q} \geq \pmb{0}$, the following properties are satisfied.
1) Positivity: $\pmb{D}(\pmb{q}) > \pmb{0}$.
2) Monotonicity: If $\pmb{q}^{(1)} \geq \pmb{q}^{(2)}$, then $\pmb{D}(\pmb{q}^{(1)}) \geq \pmb{D}(\pmb{q}^{(2)})$.
3) Scalability: For all $\lambda >1$, $\lambda \pmb{D}(\pmb{q}) >\pmb{D}(\lambda \pmb{q})$.}
Letting $\pmb{f}=[f_{11}, \cdots, f_{IM}]^T$, we have the following theorem.
\begin{theorem}\label{theorempower1}
$\pmb{f}(\pmb{q})$ is a standard interference function.
\end{theorem}

\itshape \textbf{Proof:}  \upshape
Please refer to Appendix B.
 \hfill $\Box$
 
Based on Theorem 2, we have the following corollaries.

\begin{coro}
If there exists $\pmb q$ such that $\pmb q \geq \pmb f(\pmb q)$, the iterative fixed-point method $\pmb q^{(t+1)}=\pmb f(\pmb q^{(t)})$
will converge to the unique fixed point $\pmb q^* =\pmb f(\pmb q^*)$ with any initial point $\pmb q^{(0)}$.
\end{coro}

\itshape \textbf{Proof:}  \upshape
Please refer to \cite[Theorem~2]{yates1995framework}.
 \hfill $\Box$

\begin{coro}
If Problem (\ref{max2}) is feasible, the optimal $\pmb q^*$ of Problem (\ref{max2}) is component-wise minimum in the sense that any other feasible solution $\pmb q'$ that meets constraints (\ref{max2}b)-(\ref{max2}c) and $\pmb q'\geq \pmb 0$ satisfying $\pmb q' \geq \pmb q^*$.
\end{coro}

\begin{coro}
Problem (\ref{max2}) is feasible, if and only if there exists $\pmb q^*=\pmb f(\pmb q^*)$ and $\sum_{m=1}^M q_{im}^*\leq Q_i, \forall i \in \mathcal I$.
If Problem (\ref{max2}) is feasible, the optimal solution $\pmb q^*$ to Problem (\ref{max2}) is unique with satisfying $\pmb q^*=\pmb f(\pmb q^*)$.
\end{coro}

Since Corollary 2 and 3 can be easily proved by using the same method in \cite[Theorem~2]{yates1995framework}, the proofs are omitted.
From Corollary 1 to Corollary 3, the convergence and global optimality of DPC-SPM algorithm can be verified.
\subsection{Further Discussion}
Considering the special case where users in the same cell on the same subchannel are with equal rate demands, i.e., $R_{i(J_{i(m-1)}+1)}=\cdots=R_{iJ_{im}}=R_{im}$, we show that user with poor channel should be allocated with more power than user with better channel in the same cell, i.e., $ p_{i(J_{i(m-1)}+1)m}>\cdots> p_{iJ_{im}m}$.
From (\ref{power_mini2}), for all $J_{i(m-1)}+2 \leq j\leq J_{im}$, it is verified that:
\begin{equation}\label{power_mini2_2_2_2}
\begin{aligned}
 &p_{i(j-1)m}- p_{ijm}
=
\sum_{n=j-1}^{J_{im}}
  \left(2^{\frac{R_{im}}B} -1\right)2^{\frac{(n-j+1)R_{im}}B} H_{inm}
\\&\quad
  +\sum_{n=j+1}^{J_{im}}
  \left(2^{\frac{R_{im}}B} -1\right) 2^{\frac{(n-j-1)R_{im}}B}H_{inm}
  \\&
  \quad-2\sum_{n=j}^{J_{im}}
  \left(2^{\frac{R_{im}}B} -1\right)2^{\frac{(n-j)R_{im}}B}H_{inm}
 \\
  & =\! \sum_{n=j}^{J_{im}}
  \left(2^{\frac{R_{im}}B} \!-\!1\right) 2^{\frac{(n-j+1)R_{im}}B}H_{inm}\!+\!
  \left(2^{\frac{R_{im}}B} \!-\!1\right)H_{i(j-1)m}
  \\
  & \quad\! -\!
  \!\left(\!2^{\frac{R_{im}}B} \!-\!1\!\right)\!\! 2^{\frac{-R_{im}}B} \!  H_{ijm}
\!+\!
  \sum_{n=j}^{J_{im}}
  \!\left(\!2^{\frac{R_{im}}B}\! -\!1\!\right)\! \!2^{\frac{(n-j-1)R_{im}}B} \!H_{inm}
  \\&\quad -2\sum_{n=j}^{J_{im}}
  \left(2^{\frac{R_{im}}B} -1\right) 2^{\frac{(n-j)R_{im}}B}H_{inm}
  \\
  &
  >
  \sum_{n=j}^{J_{im}}{
  \left(2^{\frac{R_{im}}B} -1\right) 2^{\frac{(n-j)R_{im}}B}}
  \left( 2^{\frac{R_{im}}B}
  +2^{\frac{-R_{im}}B}
  -2
  \right)H_{inm}
    \\
  &
  >0,\nonumber
\end{aligned}
\end{equation}
where the first inequality follows from
\begin{eqnarray}\label{power_min2_22}
&&\!\!\!\!\!\!\!\!\!\!\!H_{i(j-1)m}
=\max_{l\in\{j-1, \cdots, J_{im}\}}\frac{\sum_{k \in \mathcal I \setminus \{i\}} q_{km} |h_{klm}|^2+\sigma^2}{|h_{ilm}|^2}
\nonumber \\
&&\!\!\!\!\!\!\!\!\!\!\!\quad \geq\max_{l\in\{j, \cdots, J_{im}\}}\frac{\sum_{k \in \mathcal I \setminus \{i\}} q_{km } |h_{klm}|^2+\sigma^2}{|h_{ilm}|^2}= H_{ijm}
\end{eqnarray}
and $2^{\frac{R_{im}}B}> 1$,
and the second inequality holds because $x+x^{-1}>2$ for $x>1$ and $0<x<1$.

To implement the NOMA scheme, each BS broadcasts the channel gain orders and the cancellation schemes to all served users.
The weakest user directly decodes its own message.
With the help of SIC, the strong user decodes the messages in two stages.
In the first stage, the strong user needs to decode the messages of weak users served by the same BS on the same subchannel.
In the next stage, the strong user subtracts the decoded messages and then decodes its own message.
For OFDMA and broadcast channel (BC), each user only needs to decode its own message, i.e., all users do not need to conduct SIC.
Thus, for NOMA, the BSs need to broadcast additional information to assist SIC of the users and the receivers are complicated compared with OFDMA and BC.
\section{Sum Rate Maximization for Multi-Cell Networks}
In this section,  we solve the sum rate maximization Problem (\ref{max1}) with
\begin{equation}\label{sumratemaxobj2}
V(\pmb p, \pmb q)=-\sum_{i =1}^I \sum_{m=1}^M \sum_{j=J_{i(m-1)}+1}^{J_{im}}r_{ijm}.
\end{equation}
For sum rate maximization Problem (\ref{max1}), we show that it can be decoupled into two subproblems, i.e., power allocation problem in a single cell, and power control problem in multiple cells.
Given total transmission power $\pmb q$,
inter-cell interference can be evaluated as constant value, hence the
sum rate maximization Problem (\ref{max1}) can be further decoupled into multiple single-cell power allocation problems.
The power allocation problem in a single cell can be proved to be convex by checking the convexity of the objective function.
By solving the KKT conditions, the closed-form expression of power allocation problem in a single cell can be obtained.
Based on the results of power allocation problem in a single cell, the original sum rate maximization Problem (\ref{max1}) can be transformed into an equivalent problem with fewer variables.
A distributed algorithm is proposed to solve the transformed problem.

With given total transmission power $\pmb q$, sum rate maximization Problem (\ref{max1}) with objective function (\ref{sumratemaxobj2}) becomes the following problem,
\begin{subequations}\label{ratemax2}
\begin{align}
\mathop{\min}_{\pmb{p}\geq\pmb 0}\;
  &  - \sum_{i =1}^I\!\sum_{m=1}^M \!\sum_{j=J_{i(m-1)}+1}^{J_{im}}\!
B\log_2\left(\!
1\!+\!\frac{ p_{ijm}}
{  \sum\limits_{n=j+1}^{J_{im}}\! p_{inm}\!+\!
H_{ijm}}
\!\right)
\\
\textrm{s.t.}\quad   \!\!
&
\sum_{j=J_{i(m-1)}+1}^{J_{im}}p_{ijm}=q_{im}, \quad \forall i \in \mathcal I, m \in \mathcal M\\
&
 p_{ijm}
 \geq \left(2^{\frac{R_{ijm}}B}-1\right) \left(
 \sum_{n=j+1}^{J_{im}} p_{inm}+
H_{ijm}
 \right),
 \nonumber \\ & \qquad\forall i \in \mathcal I, m \in \mathcal M, j \in \mathcal J_{im}.
\end{align}
\end{subequations}

Since Problem (\ref{ratemax2}) has a decoupling objective function and  decoupling constraints, Problem (\ref{ratemax2}) can be decoupled into $IM$ individual problems.
Having solved the $IM$ individual problems, we can substitute those $IM$ optimal solutions into Problem (\ref{max1}), which results in the power control problem in multiple cells.
Thus, the original Problem (\ref{max1}) can be decoupled into power allocation problem in a single cell and power control problem in multiple cells.

\subsection{Power Allocation in A Single Cell}
For Problem (\ref{ratemax2}), power allocation problem for BS $i$ on subchannel $m$ is formulated as
\begin{subequations}\label{ratemax3}
\begin{align}
\mathop{\min}_{\pmb{p}_{im}\geq \pmb 0}\;
  &  -\sum_{j=J_{i(m-1)}+1}^{J_{im}}
 B\log_2\left(
1+\frac{ p_{ijm}}
{ \sum_{n=j+1}^{J_{im}} p_{inm}+
H_{ijm}}
\right)
\\
\textrm{s.t.}\quad
&
\sum_{j=J_{i(m-1)}+1}^{J_{im}}p_{ijm}=q_{im}  \\
&
 p_{ijm}
 \geq \left(2^{\frac{R_{ijm}}B} -1\right)\left(
 \sum_{n=j+1}^{J_{im}} p_{inm}+
H_{ijm}
 \right),
 \nonumber \\& \qquad  \forall j\in \mathcal J_{im},
\end{align}
\end{subequations}
where $\pmb p_{im}=[p_{i(J_{i(m-1)}+1)m}, \cdots, p_{iJ_{im}m}]^T$.

Since the convexity of objective function (\ref{ratemax3}a) cannot be easily checked, the sum rate maximization problem was regarded as a nonconvex multivariate optimization problem via nonlinear programming approaches \cite{zhang2016noma,7587811}.
To solve Problem (\ref{ratemax3}), we show that Problem (\ref{ratemax3}) is indeed convex by checking the convexity of objective function (\ref{ratemax3}a).
\begin{theorem}
Problem (\ref{ratemax3}) is a convex problem, and there exists at least one feasible solution to Problem (\ref{ratemax3}) if and only if the total transmission power of BS $i$ and rate constraints of users served by BS $i$ satisfy the following condition:
\begin{equation}\label{ratemaxfeasiblecond}
\sum_{j=J_{i(m-1)}+1}^{J_{im}}
{ \left(2^{\frac{R_{ijm}}B} -1\right)}
{2^{\sum_{s=J_{i(m-1)}+1}^{j-1}\frac{R_{ism}}B}}H_{ijm}\leq q_{im}.
\end{equation}
\end{theorem}

\itshape \textbf{Proof:}  \upshape
Please refer to Appendix C.
 \hfill $\Box$

Due to that Problem (\ref{ratemax3}) is convex, we can obtain the globally optimal solution by solving the KKT conditions \cite{boyd2004convex,bertsekas2009convex}, and the following theorem is hence provided.

\begin{theorem}
The optimal power allocation strategy for each BS is to allocate additional power to the user with the best channel gain, while other users served by this BS are allocated with minimal power to maintain their minimal rate demands.
More specifically, the globally optimal power allocation for Problem (\ref{ratemax3}) equals
\begin{equation}\label{ratemaxeq1}
\begin{aligned}
& \!\!\!\!\! p_{ijm} \!\!=\!\!
  \begin{cases}
\frac{q_{im}\!\left(2^{\frac{R_{ijm}}B} \!-\!1\right)}{ 2^{\sum_{n\!=\!J_{i(m-1)}\!+\!1}^{j} \!\frac{R_{inm}}B}}\!+\!\frac{\left(\!2^{\frac{R_{ijm}}B}\!-\!1\!\right)\!H_{ijm}}{  2^{\frac{R_{ijm}}B}}
  -  \\
\sum_{n=J_{i(m-1)}+1}^{j-1}\!\frac{\left(\!2^{\frac{R_{ijm}}B}\!-\!1\!\right)\!\!\left(\!2^{\frac{R_{inm}}B}\!-\!1\!\right)\!H_{inm}}
{  2^{\sum_{l=n}^{j}\frac{R_{ilm}}B}},  &\!\!\!\! \!\!\mbox{ if $j<J_{im}$} \\
\frac{q_{im}}{ 2^{\sum_{n=J_{i(m-1)}+1}^{J_{im}-1} \frac{R_{inm}}B}}
-  \\
\sum_{n=J_{i(m-1)}+1}^{J_{im}-1}\frac{\left(2^{\frac{R_{inm}}B}-1\right)H_{inm}}{ 2^{\sum_{l=n}^{J_{im}-1} \frac{R_{ilm}}B}}, &\!\! \!\!\!\!
\mbox{ if $j=J_{im}$}
  \end{cases},\!\!\!\!\!
\end{aligned}
\end{equation}
and the corresponding optimal value of Problem (\ref{ratemax3}) is
\begin{eqnarray}\label{ratemaxeq2}
-B
\log_2&&\!\!\!\!\!\!\!\!\!\!\!\left( 1 + \frac{q_{im} }{ 2^{\sum_{j=J_{i(m-1)}+1}^{J_{im}-1}\frac{R_{ijm}}B}H_{i J_{im}m}}
 - \sum_{j=J_{i(m-1)}+1}^{J_{im}-1} \right.
 \nonumber\\&&\!\!\!\!\!\!\!\!\!\!\!
 \left.
\frac{ \left(2^{\frac{R_{ijm}}B}-1\right)H_{ijm}}
 {
 2^{\sum_{l= j}^{J_{im}-1}\frac{R_{ilm}}B}H_{i J_{im}m}} \right)
-
\sum_{j=J_{i(m-1)}+1}^{J_{im}-1}R_{ijm},
\end{eqnarray}
where the first term is the negative rate of user $J_{im}$ with the largest channel gain among users served by BS $i$ on subchannel $m$, and the second term is the negative sum rate of all other users served by BS $i$ on subchannel $m$.
\end{theorem}

\itshape \textbf{Proof:}  \upshape
Please refer to Appendix D.
 \hfill $\Box$

Similar to (\ref{power_mini2}), the concept of strong user reflected from (\ref{sys2_2}), (\ref{sys3}) and (\ref{sys3_2}) is helpful in obtaining the closed-form expression in (\ref{ratemaxeq1}).
Especially, the decreasing order of $H_{ijm}$ in (\ref{power_min2_22}) obtained from (\ref{sys3_2}) is crucial in proving that Problem (\ref{ratemax3}) is convex, which fortunately has closed-form solution.
According to (\ref{ratemaxeq2}), we can find that the optimal sum rate is mainly determined by the rate of user $J_{im}$ with best channel gain among users served by BS $i$ on subchannel $m$ and the optimal sum rate increases with the total transmission power of BS $i$ on subchannel $m$ on a logarithmic scale.
Note that we obtain the similar conclusion as in \cite{7890454}.
The difference is that the inter-cell interference is not considered in \cite{7890454}.

\subsection{Power Control in Multiple Cells}
Based on Theorem 3 and Theorem 4, we can readily transform sum rate maximization Problem (\ref{max1})  with objective function (\ref{sumratemaxobj2})  into an equivalent problem.
\begin{theorem}
Sum rate maximization Problem (\ref{max1})  with objective function (\ref{sumratemaxobj2})  is equivalent to the following problem:
\begin{subequations}\label{ratemaxPC1}
\begin{align}
\mathop{\min}_{\pmb q \geq \pmb 0, \pmb x\geq \pmb 0}\;
 \quad& \! -\!\sum_{i=1}^I \sum_{m=1}^M B \left(\! 1 \!+\! \frac{q_{im} }{ 2^{\sum_{j=J_{i(m-1)}+1}^{J_{im}-1}\frac{R_{ijm}}B}x_{i J_{im}m}}
\right.
 \nonumber\\&
 \left.
\quad -  \sum_{j=J_{i(m-1)}+1}^{J_{im}-1}\frac{ \left(2^{\frac{R_{ijm}}B}-1\right)x_{ijm}}
 {
 2^{\sum_{l= j}^{J_{im}-1}\frac{R_{ilm}}B}x_{i J_{im}m}} \!\right)\\
\textrm{s.t.}\quad\qquad \!\!\!\!\!\!\!\!\!
&
q_{im}\geq \sum_{j=J_{i(m-1)}+1}^{J_{im}}
{ \left(2^{\frac{R_{ijm}}B} -1\right)}
\nonumber \\ &\qquad{2^{\sum_{s=J_{i(m-1)}+1}^{j-1}\frac{R_{ism}}B}} x_{ijm}, \quad \forall i\in \mathcal I, m \in \mathcal M \\
&x_{ijm} \geq  \frac{\sum_{k \in \mathcal I \setminus \{i\}} q_{km} |h_{klm}|^2 + \sigma^2}{|h_{ilm}|^2},
 \nonumber\\ & \qquad \forall i \in \mathcal I ,m  \in  \mathcal M,j \in \mathcal J_{im},  {l\in\{j, \cdots, J_{im}\}}  \\
&\sum_{m=1}^M q_{im}\leq Q_i, \quad \forall i \in \mathcal I,
\end{align}
\end{subequations}
where $\pmb x=[x_{111}, \cdots, x_{1J_{1M}{M}}, \cdots, x_{IJ_{IM}M}]^T$
and the power for each user is given by (\ref{ratemaxeq1}).
\end{theorem}

Note that there always exists $l\in\{j, \cdots, J_{im}\}$ such that $x_{ijm}= \frac{\sum_{k \in \mathcal I \setminus \{i\}} q_{km} |h_{klm}|^2+\sigma^2}{|h_{ilm}|^2}$ for all $i \in \mathcal I, m \in \mathcal M, j\in\mathcal J_{im}$ as the objective function increases with $x_{ijm}$, which shows that $x_{ijm}=H_{ijm}$ defined in (\ref{sys3_2}).
Since Theorem 5 can be easily proved through substituting (\ref{ratemaxeq1}) into Problem (\ref{max1}) with objective function (\ref{sumratemaxobj2}), the proof of Theorem 5 is omitted.
It can be found that the objective function in Problem (\ref{ratemaxPC1}) is still nonconvex and the constraints of Problem (\ref{ratemaxPC1}) are all linear with respect to $\pmb q$ and $\pmb x$.
In the following, one distributed algorithm with low complexity is proposed.
Denote $\pmb q_i=[q_{i1}, \cdots, q_{iM}]^T$ and $\pmb x_{i}=[x_{i(J_{i0}+1)1}, \cdots, x_{iJ_{iM}M}]^T$.
With given $\pmb q_{-i}=[\pmb q_1^T, \cdots, \pmb q_{i-1}^T, \pmb q_{i+1}^T, \cdots, \pmb q_{I}^T]^T$ and $\pmb x_{-i}=[\pmb x_1^T, \cdots, \pmb x_{i-1}^T, \pmb x_{i+1}^T, \cdots, \pmb x_{I}^T]^T$, Problem (\ref{ratemaxPC1}) becomes the following optimization problem for BS $i$,
\begin{subequations}\label{ratemaxPC1_11}
\begin{align}
\mathop{\min}_{\pmb q_i \geq \pmb 0, \pmb x_i\geq \pmb 0}\;
 \quad&  -\sum_{m=1}^M B \left( 1 + \frac{q_{im} }{ 2^{\sum_{j=J_{i(m-1)}+1}^{J_{im}-1}\frac{R_{ijm}}B}x_{i J_{im}m}}
\right.
 \nonumber\\&
 \left.
 \quad -  \sum_{j=J_{i(m-1)}+1}^{J_{im}-1}\frac{ \left(2^{\frac{R_{ijm}}B}-1\right)x_{ijm}}
 {
 2^{\sum_{l= j}^{J_{im}-1}\frac{R_{ilm}}B}x_{i J_{im}m}} \right)\\
\textrm{s.t.}\quad\qquad \!\!\!\!\!\!\!\!\!
&
q_{im}\geq \sum_{j=J_{i(m-1)}+1}^{J_{im}}
{ \left(2^{\frac{R_{ijm}}B} -1\right)}
\nonumber \\
& \qquad {2^{\sum_{s=J_{i(m-1)}+1}^{j-1}\frac{R_{ism}}B}}x_{ijm}, \quad m \in \mathcal M \\
&x_{ijm}\geq H_{ijm}, \quad \forall m \in \mathcal M, j\in\mathcal J_{im}     \\
&q_{im}\leq Q_{im},\quad \forall m \in \mathcal M\\
&\sum_{m=1}^M q_{im}\leq Q_i,
\end{align}
\end{subequations}
where $H_{ijm}$ is defined in (\ref{sys3_2}),
and constraints (\ref{ratemaxPC1_11}d) follow from (\ref{ratemaxPC1}c) with
\begin{eqnarray}\label{sumrateDPCMessageQim}
Q_{im}=&&\!\!\!\!\!\!\!\!\!\!\! \min_{n\in\mathcal N\setminus\{i\}, j \in \mathcal J_{nm}, l\in\{j, \cdots, J_{nm}\}}
\nonumber \\
&&\!\!\!\!\!\!\!\!\!\!\!
\frac{|h_{nlm}|^2x_{njm}-\sum_{k\in\mathcal \setminus\{i,n\}}q_{km}|h_{klm}|^2-\sigma^2}
{|h_{ilm}|^2}.
\end{eqnarray}

The representation of (\ref{ratemaxPC1_11}a) is similar to the DC problem representation \cite{5519540,7063568,7523951}.
Thus, objective function (\ref{ratemaxPC1_11}a) can be rewritten as
\begin{eqnarray*}\label{ratemaxPCeq2}
&&\!\!\!\!\!\!\!\!\!\!\!\!
   \!-\!\sum_{m=1}^M\!B  \log_2 \! \left( \! x_{iJ_{im}m}  \!+ \! \frac{q_{im} }{ 2^{\sum_{j=J_{i(m-1)}+1}^{J_{im}-1}\frac{R_{ijm}}B}}
 \right.\nonumber\\
&&\!\!\!\!\!\!\!\!\!\!\!\!
\left.
-
  \sum_{j=J_{i(m-1)}+1}^{J_{im}-1}\frac{  \left( 2^{\frac{R_{ijm}}B} - 1 \right)x_{ijm}}
 {
 2^{\sum_{l=j}^{J_{im}-1}\frac{R_{ilm}}B}} \right)
 + \sum_{m=1}^M B\log_2(x_{iJ_{im}m}) .
\end{eqnarray*}

Letting
\begin{eqnarray*}\label{ratemaxPCeq4}
F_i(\pmb q_i,\pmb x_i)  = &&\!\!\!\!\!\!\!\!\!\!\!\!- \sum_{m=1}^M B  \log_2  \left(  x_{iJ_{im}m}  +  \frac{q_{im} }{ 2^{\sum_{j=J_{i(m-1)}+1}^{J_{im}-1}\frac{R_{ijm}}B}}
 \right.\nonumber\\
&&\!\!\!\!\!\!\!\!\!\!\!\!\left.
-
  \sum_{j=J_{i(m-1)}+1}^{J_{im}-1}\frac{  \left( 2^{\frac{R_{ijm}}B} - 1 \right)x_{ijm}}
 {
 2^{\sum_{l=j}^{J_{im}-1}\frac{R_{ilm}}B}} \!\right),
\end{eqnarray*}
and
\begin{equation*}\label{ratemaxPCeq3}
G_i(\pmb x_i)=-\sum_{m=1}^M  B\log_2(x_{iJ_{im}m}),
\end{equation*}
Problem (\ref{ratemaxPC1_11}) can be written as
\begin{subequations}\label{ratemaxPC2}
\begin{align}
\mathop{\min}_{\pmb q_i\geq \pmb 0, \pmb x_i\geq \pmb 0}\;
 \quad&  K_i(\pmb q_i, \pmb x_i)= F_i(\pmb q_i, \pmb x_i)- G_i(  \pmb x_i)\\
\textrm{s.t.}\quad\qquad \!\!\!\!\!\!\!\!\!
&
 (\ref{ratemaxPC1_11}b), (\ref{ratemaxPC1_11}c), (\ref{ratemaxPC1_11}d), (\ref{ratemaxPC1_11}e).
\end{align}
\end{subequations}

Define function $v(y,z)=-\log_2(y-z)$.
Since
\begin{eqnarray}
\bigtriangledown^2 v
&&\!\!\!\!\!\!\!\!\!\!
=
\begin{pmatrix}
\frac{\partial ^2 v}{\partial y^2}&\frac{\partial ^2 v}{\partial y\partial z} \\
\frac{\partial ^2 v}{\partial y \partial z}&\frac{\partial ^2 v}{\partial z^2}
\end{pmatrix}
\nonumber\\
&&\!\!\!\!\!\!\!\!\!\!
=\frac{1}{(\ln2)(y-z)^2}
\begin{pmatrix}
1&-1 \\
-1&1
\end{pmatrix}
\nonumber\\
&&\!\!\!\!\!\!\!\!\!\!
=\frac{1}{(\ln2)(y-z)^2}
(1,-1)^T (1, -1)
\succeq \pmb 0,\nonumber
\end{eqnarray}
both $F_i(\pmb q_i, \pmb x_i)$ and $G_i(\pmb x_i)$ can be proved convex according to the nonnegative weighted sums operation and composition operation with an affine mapping that preserve convexity \cite[Page~79]{boyd2004convex}.
Therefore, the DC programming approach can be applied to realize multi-cell power control \cite{5519540}.
From \cite{5519540}, nonconvex Problem (\ref{ratemaxPC2}) can be solved suboptimally by 
converting a nonconvex problem to convex subproblems with replacing the term $-G_i(\pmb x_i)$ in the objective function (\ref{ratemaxPC2}a) with its convex majorant $-G_i(\tilde{\pmb x}_i)-\triangledown G_i^T(\tilde{\pmb x}_i)(\pmb x_i -\tilde{\pmb x}_i)$.
$\triangledown G_i(\tilde{\pmb x}_i)$ is the gradient of $G_i(\pmb x_i)$ at point $\tilde{\pmb x}_i$, and $\triangledown G_i(\tilde{\pmb x}_i)$ can be calculated by
\begin{eqnarray}\label{ratemaxPCeq5}
\triangledown G_i(\tilde {\pmb x}_i)=
\frac {B} {\ln 2}
\left[0, \cdots, - \frac{1}{\tilde x_{iJ_{i1}1}},
\cdots,
 -\frac{1}{\tilde x_{iJ_{iM}M}}
\right]^T.
\end{eqnarray}
The convex optimization problem in (\ref{ratemaxPC3}) can be solved by the interior point method \cite{boyd2004convex,bertsekas2009convex}.

In the following, we provide a distributed power control algorithm to solve sum rate maximization Problem (\ref{ratemaxPC1}) in Algorithm 2.

\begin{algorithm}[h]
\caption{Distributed Power Control for Sum Rate Maximization (DPC-SRM)}
\begin{algorithmic}[1]
\State Initialize $\pmb q^{(0)}$, $\pmb x^{(0)}$,  the iteration number $t=0$, and the tolerance $\epsilon$.
\For{$i=1, 2, \cdots, I$}
\State $\tilde{\pmb x}_i=\pmb x_i^{(t)}$.
\Repeat
\State Define a convex approximate  function of $\pmb H_i(\pmb x_i)$ at point $\tilde {\pmb x}_i$ as
\begin{equation}\label{ratemaxPCeq6}
\tilde {K}_i(\pmb q_i, \pmb x_i)=F_i(\pmb q_i, \pmb x_i)-G_i(\tilde{\pmb x}_i)-\triangledown G^T(\tilde{\pmb x}_i)(\pmb x_i -\tilde{\pmb x}_i).
\end{equation}
\State
Solve the convex optimization problem
\begin{equation}\label{ratemaxPC3}
(\tilde{\pmb q}_i,\tilde{\pmb x}_i)=\left\{ \begin{array}{ll}
\!\!\arg\mathop{\min}\limits_{\pmb q_i\geq \pmb 0, \pmb x_i\geq \pmb 0}
&\!\!
\tilde K_i(\pmb q_i, \pmb x_i)
\\
\quad\textrm{s.t.}
&\!\! (\ref{ratemaxPC1_11}b), (\ref{ratemaxPC1_11}c), (\ref{ratemaxPC1_11}d), (\ref{ratemaxPC1_11}e).
\end{array} \right.
\end{equation}
\Until the objective function (\ref{ratemaxPC2}a) converges
\State $q_i^{(t+1)}=\tilde{\pmb q}_i$, $\pmb x_i^{(t+t)}=\tilde{\pmb x}_i$.
\EndFor
\State
If $|\sum_{i=1}^I  K_i(\pmb q_i^{(t)},\pmb x_i^{(t)})-\sum_{i=1}^I  K_i(\pmb q_i^{(t+1)},\pmb x_i^{(t+1)})|\leq \epsilon$, terminate.
Otherwise, set $t=t+1$ and go to step 2.
\end{algorithmic}
\end{algorithm}
\subsection{Convergence and Complexity Analysis}
According to \cite{5519540}, the DC programming steps (i.e., Step 4 to Step 7) in DPC-SRM algorithm always converge to a suboptimal stationary point, i.e., $K_i(\pmb q_i^{(t)},\pmb x_i^{(t)})\geq K_i(\pmb q_i^{(t+1)},\pmb x_i^{(t+1)})$, which shows that the objective value (\ref{ratemaxPC1}a) is nonincreasing when sequence ($\pmb q, \pmb x$) is updated.
Furthermore, the objective value (\ref{ratemaxPC1}a) can be found upper-bounded.
As a result, DPC-SRM algorithm must converge.

For DPC-SRM algorithm, the major complexity in each iteration lies in solving convex optimization problem  (\ref{ratemaxPC3}).
Assume that the number of users on every subchannel in each cell is $N$.
Considering that the dimension of variables in problem (\ref{ratemaxPC3}) is $M(N+1)$,
the complexity of solving problem (\ref{ratemaxPC3}) by using the standard
interior point method is $\mathcal O(M^3 N^3)$ \cite[Page 487, 569]{boyd2004convex}.
Hence, the total complexity of DPC-SRM is $\mathcal O(L_{\text{SR}}L_{\text{DC}}IM^3N^3)$, where $L_{\text{SR}}$ and $L_{\text {DC}}$ denote the total number of iterations of the out layer of DPC-SR algorithm and the DC programming, respectively.
\subsection{Implementation Method}
To implement the proposed DPC-SRM algorithm, each BS $i$ needs to update power vector $\pmb q_i$ and auxiliary vector $\pmb {x}_i$ by solving Problem (\ref{ratemaxPC1_11}).
Solving Problem (\ref{ratemaxPC1_11}) involves $R_{ijm}$, $H_{ijm}$, $Q_{im}$, $\forall m \in \mathcal M$, $j \in \mathcal J_{im}$.
Assume that the minimal rate demands for users served by BS $i$ are available at BS $i$.
Since $H_{ijm}=\max_{l\in\{j, \cdots, J_{im}\}}\frac{\sum_{k \in \mathcal I \setminus \{i\}} q_{km} |h_{klm}|^2+\sigma^2}{|h_{ilm}|^2}$,
numerator $|h_{ilm}|$ is the channel gain between BS $i$ and its served user $l$ on subchannel $m$,
which can be estimated by channel reciprocity.
Beside,
$\sum_{k \in \mathcal I \setminus \{i\}} q_{km} |h_{klm}|^2+\sigma^2$ is the total interference power of user $l$ served by BS $i$ on subchannel $m$, and the value of interference power can be measured by user $l$.
Due to the fact that
\begin{eqnarray*}
Q_{im}&&\!\!\!\!\!\!\!\!\!\!\!\!=\min_{n\in\mathcal N\setminus\{i\}, j \in \mathcal J_{nm}, l\in\{j, \cdots, J_{nm}\}}
\nonumber \\
&&\!\!\!\!\!\!\!\!\!\!\!\!\frac{|h_{nlm}|^2x_{njm}-\sum_{k\in\mathcal \setminus\{n\}}q_{km}|h_{klm}|^2-\sigma^2+q_{im}|h_{ilm}|^2}
{|h_{ilm}|^2},
\end{eqnarray*}
numerator $|h_{ilm}|$ is the cross channel gain between BS $i$ and user $l$ served by BS $n$ $(n\neq i)$ on subchannel $m$, which can be estimated at BS $i$ for receiving the pilot from user $l$ according to channel reciprocity.
The denominator of calculating $Q_{im}$ contains three parts.
In the first part, i.e., $|h_{nlm}|^2x_{njm}$,  $|h_{nlm}|$ is the channel gain between BS $n$ and its served user $l$ on subchannel $m$ and $x_{njm}$ is the strategy of BS $n$.
Hence, $|h_{nlm}|^2x_{njm}$ is known at BS $n$.
In the second part, $\sum_{k\in\mathcal \setminus\{n\}}q_{km}|h_{klm}|^2+\sigma^2$ is the total interference power of user $l$ served by BS $n$ on subchannel $m$.
In the third part $q_{im}|h_{ilm}|^2$, $q_{im}$ is the transmission power of BS $i$ on subchannel $m$.
To calculate $Q_{im}$, user $l$ served by BS $n$ sends its overall received interference and noise to BS $n$.
Then, having obtained the messages from its served users, BS $n$ calculates $|h_{nlm}|^2x_{njm}-\sum_{k\in\mathcal \setminus\{n\}}q_{km}|h_{klm}|^2-\sigma^2$ and sends these calculated values to BS $i$, which helps BS $i$ calculate $Q_{im}$.
As a result, BS $i$ calculates the optimal $\pmb q_i$ and $\pmb {x}_i$ by solving Problem (\ref{ratemaxPC1_11}).
Each BS updates its power vector and auxiliary vector until the total interference power of each user converges.

\section{Extension to MIMO-NOMA Systems}
Consider a downlink multi-cell network with NOMA, 
where there are $I$ BSs with $M$ antennas each and $J$ users with $N$ antennas each.
For MIMO-NOMA, superposition coding is employed at each BS.
The transmit signal at BS $i$ is given by \cite{7236924}
\begin{equation}\label{rev2sys1nomaeq2}
{\pmb s}_i=\left[
         \begin{array}{c}
           \sqrt{p_{i(J_{i-1}+1)1}}s_{i(J_{i-1}+1)1}+\cdots+\sqrt{p_{iJ_i1}} s_{iJ_i1}\\
           \vdots \\
           \sqrt{p_{i(J_{i-1}+1)M}} s_{i(J_{i-1}+1)M} + \cdots + \sqrt{p_{iJ_iM}} s_{iJ_iM} \\
         \end{array}
       \right]
,
\end{equation}
where $s_{ijm}$ denotes the information bearing signal to be transmitted to user $j$ in cluster $m$ served by BS $i$, and $p_{ijm}$ is the NOMA power allocation coefficient.
Obviously, $p_{ijm}>0$ shows that user $j$ is assigned in cluster $m$.
It is assumed that all users are already clustered and each user is only assigned to one cluster.
Let $\mathcal M=\{1, 2, \cdots, M\}$ be the set of clusters.
The set of users in cluster $m$ is denoted by $\mathcal J_{im}=\{J_{i(m-1)}+1, J_{i(m-1)}+2, \cdots, J_{im}\}$, where $J_{i0}=J_{i-1}+1$, $J_{iM}=J_i$, $J_{im}=\sum_{l=1}^m |\mathcal J_{il}|$.
The observation at user $j\in \mathcal J_{im}$ in cluster $m$ is given by
\begin{eqnarray}\label{rev2sys2}
\!\!\!\!\!\!\pmb y_{ijm}
&&\!\!\!\!\!\!\!\!\!\!\!\!
=\pmb H_{ij} \pmb s_{i}+\sum_{k \in \mathcal I \setminus \{i\}} \pmb H_{kj}  \pmb s_{k}+\pmb n_{jm}
\nonumber\\
&&\!\!\!\!\!\!\!\!\!\!\!\!
=\pmb h_{ijm} \sqrt{p_{ijm}}  s_{ijm}+
\underbrace{\sum_{l\in\mathcal J_{im}\setminus\{j\}}\pmb h_{ijm} \sqrt{p_{ilm}}  s_{ilm}}_{\text{intra-cell intra-cluster interference}}
\nonumber\\
&&\!\!\!\!\!\!\!\!\!\!\!\!
\quad  +
\underbrace{\sum_{t\in\mathcal M\setminus\{m\}}
\sum_{l=J_{i(t-1)}+1}^{J_{it}}\pmb h_{ijt} \sqrt{p_{ilt}}  s_{ilt}}_{\text{intra-cell inter-cluster interference}}
\nonumber\\
&&\!\!\!\!\!\!\!\!\!\!\!\!
\quad +
\underbrace{\sum_{k \in \mathcal I \setminus \{i\}} \sum_{t=1}^M
\sum_{n=J_{k(t-1)}+1}^{J_{kt}}\pmb h_{kjt} \sqrt{p_{knt}} s_{knt}}_{\text{inter-cell interference}} +\pmb n_{jm},
\end{eqnarray}
where $\pmb H_{ij}$ is the channel gain between BS $i$ and user $j$,
$\pmb h_{ijm}$ is the channel gain between BS $i$ and user $j$ in cluster $m$, and $\pmb n_{jm}$ represents the additive zero-mean Gaussian noise vector with variance $\sigma^2 \pmb I$.

Denote by $\pmb v_{ijm}$ the detection vector used by user $j$ served by BS $i$ in cluster $m$. After applying this vector into (\ref{rev2sys2}), the signal model can be rewritten as follows:
\begin{eqnarray}\label{rev2sys2_23}
\!\!\!\!\!\!\!\pmb v_{ijm}^H \pmb y_{ijm}
&&\!\!\!\!\!\!\!\!\!\!\!\!=\pmb v_{ijm}^H\pmb h_{ijm} \sqrt{p_{ijm}}  s_{ijm} +\pmb v_{ijm}^H\pmb n_{jm}
\nonumber\\
&&\!\!\!\!\!\!\!\!\!\!\!\!   +
\underbrace{\sum_{l \in \mathcal J_{im}\setminus\{j\}}\pmb v_{ijm}^H\pmb h_{ijm} \sqrt{p_{ilm}}  s_{ilm}}_{\text{intra-cell intra-cluster interference}}
\nonumber\\
\nonumber\\
&&\!\!\!\!\!\!\!\!\!\!\!\!  +
\underbrace{\sum_{t\in\mathcal M\setminus\{m\}}
\sum_{l=J_{i(t-1)}+1}^{J_{it}}\pmb v_{ijm}^H\pmb h_{ijt} \sqrt{p_{ilt}}  s_{ilt}}_{\text{intra-cell inter-cluster interference}}
\nonumber\\
&&\!\!\!\!\!\!\!\!\!\!\!\!  +
\underbrace{\sum_{k \in \mathcal I \setminus \{i\}} \sum_{t=1}^M
\sum_{n=J_{k(t-1)}+1}^{J_{kt}}\pmb v_{ijm}^H\pmb h_{kjt} \sqrt{p_{knt}} s_{knt}}_{\text{inter-cell interference}}.
\end{eqnarray}

The channel conditions are crucial to the implementation of NOMA.
As in \cite{7236924}, it is assumed that the channel gains are sorted as follows:
\begin{equation}\label{rev2sys2_223}
|\pmb v_{i(J_{i(m-1)}+1)m}^H \pmb h_{i (J_{i(m-1)}+1)m}|^2 \leq \cdots\leq
|\pmb v_{iJ_{im}m}^H \pmb h_{iJ_{im}m}|^2.
\end{equation}

With removing intra-cell inter-cluster interference, the detection vector can be obtained as in (13) in \cite{7236924}.
In order to remove intra-cell inter-cluster interference, the number of users' antennas is larger than or equal to that of the BS \cite{7236924}.
With detection vector $\pmb v_{ijm}$ fixed and without intra-cell inter-cluster interference,
the achievable rate of user $j\in\mathcal J_{im}$ to detect the message of user $l\in\{J_{i(m-1)}+1, \cdots, j\}$ in cluster $m$ is
\begin{equation}\label{rev2sys2_2}
r_{ijlm} =B\log_2 \left(
1 + \frac{|\pmb v_{ijm}^H \pmb h_{ijm}|^2 p_{ilm}}{|\pmb v_{ijm}^H \pmb h_{ijm}|^2 \sum_{n=l+1}^{J_{im}} p_{inm} +
 Y_{ijlm}}
 \right),
\end{equation}
where $Y_{ijlm}=
\sum_{t=1}^M \sum_{k \in \mathcal I \setminus \{i\}}  q_{kt} |\pmb v_{ijm}^H \pmb h_{kjt}|^2
+ \sigma^2 |\pmb v_{ijm}|^2$,
and
$q_{km}=\sum_{j=J_{k(m-1)}+1}^{J_{km}}p_{kjm}$ is the total transmission power of BS $k$ in cluster $m$.
According to (\ref{rev2sys2_2}), strong user $j$ with high effective channel gain needs to decode the message of weak user $l\leq j$ with low effective channel gain.
To ensure successful SIC,
the achievable rate of user $j\in\mathcal J_{im}$ in cluster $m$ can be given by
\begin{eqnarray}
&&\!\!\!\!\!\!\!r_{ijm}=\min_{l\in\{j, \cdots, J_{im}\}} r_{iljm}
\nonumber\\
&&\!\!\!\!\!\!\!
 =
\min_{l\in\{j, \cdots, J_{im}\}}B\log_2\left(
1+\frac{p_{ijm}}
{ \sum_{n=j+1}^{J_{im}} p_{inm}+
\frac{Y_{iljm}}{|\pmb v_{ilm}^H \pmb h_{ilm}|^2}}
\!\right)
\nonumber\\
&&\!\!\!\!\!\!\!=
B\log_2\left(
1+\frac{p_{ijm}}
{ \sum_{n=j+1}^{J_{im}} p_{inm}+
H_{ijm}}
\right),\label{rev2sys3}
\end{eqnarray}
where
\begin{equation}\label{rev2sys3_2}
H_{ijm}\!=\!\!\max_{l\in\{j, \cdots, J_{im}\}}\!\!\frac{\sum_{t=1}^M\!\sum_{k \in \mathcal I \setminus \{i\}} \!\!q_{kt} |\pmb v_{ilm}^H \pmb h_{klt}|^2\!+\!\sigma^2|\pmb v_{ilm}|^2}{|\pmb v_{ilm}^H \pmb h_{ilm}|^2}.
\end{equation}
Since the rate formulations (\ref{rev2sys3}) and (\ref{rev2sys3_2}) respectively have similar structures as equations (\ref{sys3}) and (\ref{sys3_2}), the power control methods in Section III and Section IV can be applied to MIMO-NOMA systems.

\section{Numerical Results}
In this section, numerical results are presented to evaluate the performance of the proposed schemes for multi-cell networks with NOMA. 
In the simulations, we consider a three-site 3GPP LTE network with an inter-site distance of 800 m, adopting a wrap-around technique \cite{Chin2015Power}.
The simulated system operates at 2 GHz, the number of subchannels is $M=10$ and the bandwidth of each subchannel is $B=1$ MHz.
The three-sector antenna pattern is used for each site and the gain for the three-sector,
of which 3dB beamwidth in degrees is 70 degrees, is 14dBi \cite{access2010further}.
In the propagation model, we use the large-scale path loss $L(d)=128.1+37.6\log(d)$, $d$ is in km, and the standard deviation of shadow fading is set as $8$ dB \cite{access2010further}.


The total number of BSs $I$ is set as 15.
To reduce the receiver complexity and error propagation due to SIC, it is reasonable for each subchannel to be multiplexed by two or three users \cite{6464495}.
In the simulations, the number of users in each cell is set as 20 and two users are paired on each subchannel.
We set the tolerance in Algorithm 1 and Algorithm 2 $\epsilon=0.001$ and noise power $\sigma^2=-114$ dBm.
We assume equal rate demands for all users (i.e., $R_{ij}=R=0.3$ Mbps, $\forall i \in \mathcal I, j \in \mathcal J_i$) and equal maximum transmission power for all BSs (i.e., $Q_i=Q$, $\forall i \in \mathcal I$).

We compare the NOMA system with two systems: the OFDMA system, where multiple users on the same subchannel are allocated with orthogonal time fractions, and the BC system, where multiple users on the same subchannel suffer from both intra-cell (without performing SIC) and inter-cell interference.
For sum power minimization, we compare the proposed power minimization scheme for NOMA systems (labeled as `NOMA-PM') by using Algorithm 1 with sum power minimization problem for OFDMA systems (labeled as `OFDMA-SP'), which can be optimally solved by using the optimal power vector algorithm in \cite[Section V]{Chin2015Power},
and sum power minimization problem for BC systems (labeled as `BC-SP'), which can be optimally solved by using the simplex method \cite{boyd2004convex}.
For sum rate maximization, we compare the proposed sum rate maximization scheme for NOMA systems (labeled as `NOMA-RM') through using Algorithm 2, with sum rate maximization problem for OFDMA systems (labeled as `OFDMA-SR'), which can be suboptimally solved by using the distributed power control and time allocation algorithm in \cite[Section V]{Yang2017Joint}, and sum rate maximization problem for BC systems (labeled as `BC-RM'), which can be suboptimally solved by using the weighted mean-square error approach \cite[Section II]{5756489}.


\begin{figure*}
\centering
\includegraphics[width=5.75in]{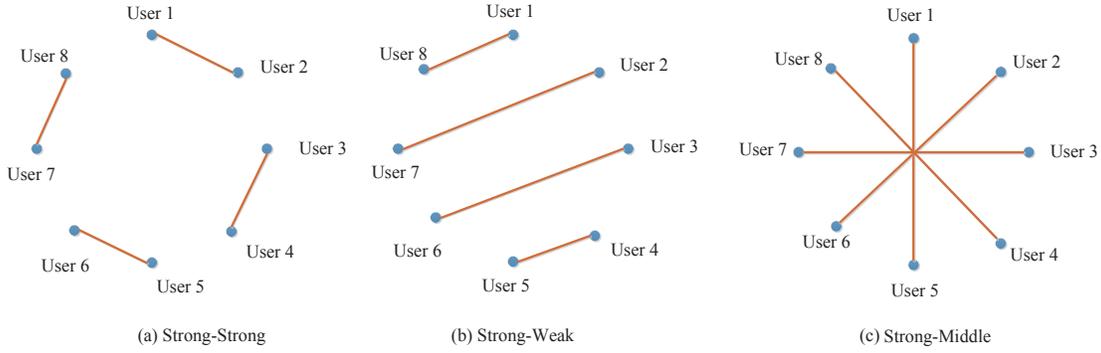}
\caption{This figure illustrates pair selection in a typical cell of 8 users.}\label{rev2userpair}
\end{figure*}

We study the influence of user pairing by considering three different user-pairing methods \cite{7273963} and \cite{lei2017}.
Fig.~\ref{rev2userpair} illustrates pair selection in a typical cell of 8 users, where users are sorted in increasing order of channel gains, i.e., user 8 enjoys the strongest channel gain while user 1 is of the weakest channel gain.
In strong-strong (SS) pair selection, the user with the strongest
channel condition is paired with the one with the second
strongest, and so on.
In strong-weak (SW) pair selection, the user with the strongest
channel condition is paired with the user with the weakest, and
the user with the second strongest is paired with one with the
second weakest, and so on.
In strong-middle (SM) pair selection, the user with the strongest
channel condition is paired with the user with the middle strongest user, i.e., user 8 is paired with user 4 in Fig.~\ref{rev2userpair}(c), and so on.

\begin{figure}
\centering
\includegraphics[width=3.0in]{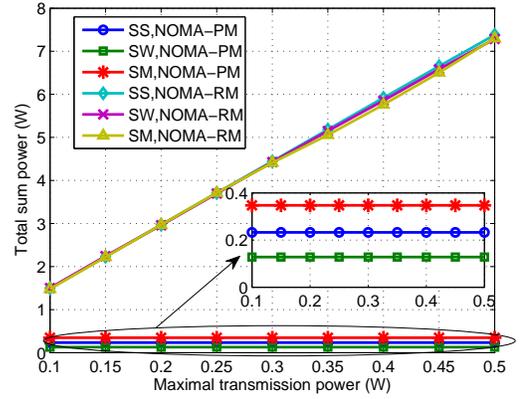}
\caption{Sum power of the system for different user-pairing methods.\label{fig11}}
\end{figure}

\begin{figure}
\centering
\includegraphics[width=3.0in]{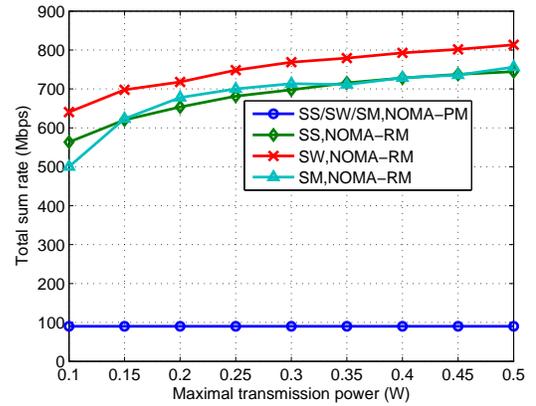}
\caption{Sum rate of the system for different user-pairing methods. \label{fig12}}
\end{figure}

In Fig. \ref{fig11} and Fig.~\ref{fig12}, we show the sum power and sum rate of the system for different user-pairing methods, respectively.
From Fig.~\ref{fig11}, it is observed that SW outperforms the other two methods in terms of power consumption for NOMA-PM.
Besides, we can also find that SW achieves the best sum rate among three user-pairing methods for NOMA-RM according to Fig.~\ref{fig12}.
Combing Fig.~\ref{fig11} and Fig.~\ref{fig12}, we can conclude that it tends to pair users with distinctive gains for both sum power minimization and sum rate maximization, which coincides with previous findings in \cite{7273963}.
Due to the superiority of SW, the following simulations are based on SW pair selection.

\begin{figure}
\centering
\includegraphics[width=3.0in]{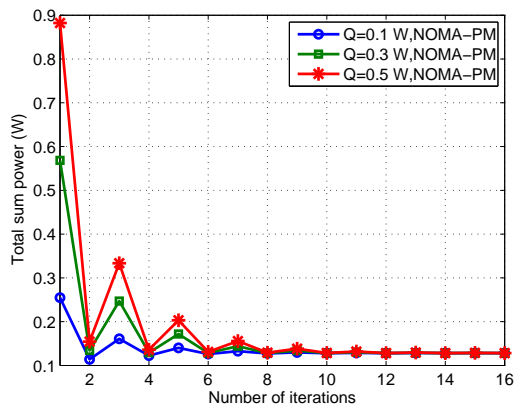}
\caption{Convergence behavior of NOMA-PM.\label{fig13}}
\end{figure}

\begin{figure}
\centering
\includegraphics[width=3.0in]{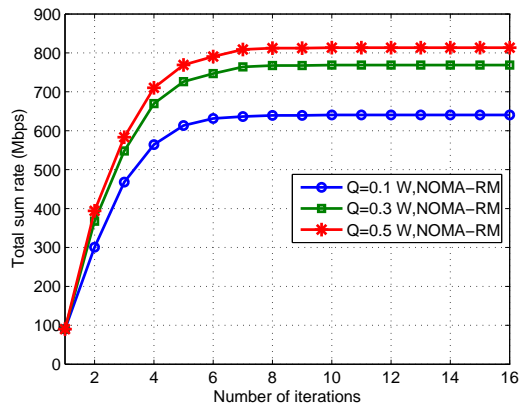}
\caption{Convergence behavior of NOMA-RM. \label{fig15}}
\end{figure}

The convergence behaviors of NOMA-PM and NOMA-RM are illustrated in Fig.~\ref{fig13} and Fig.~\ref{fig15}, respectively.
It can be seen that both NOMA-PM and NOMA-RM converge rapidly, which makes our proposed algorithms suitable for practical applications.
From Fig.~\ref{fig15}, the sum rate of NOMA-RM monotonically increases, which confirms the convergence analysis in Section IV-C.

\begin{figure}
\centering
\includegraphics[width=3.0in]{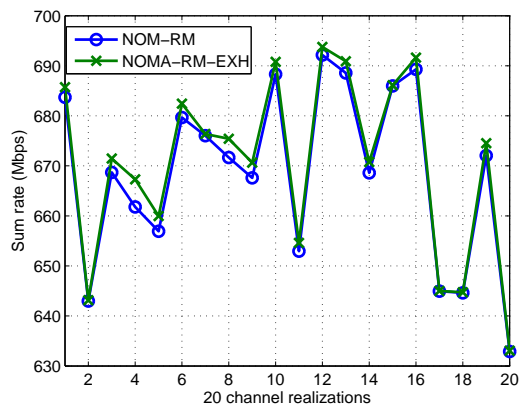}
\caption{Performance comparison of NOMA-RM and NOMA-RM-EXH.\label{fig19}}
\end{figure}

\begin{figure}
\centering
\includegraphics[width=3.0in]{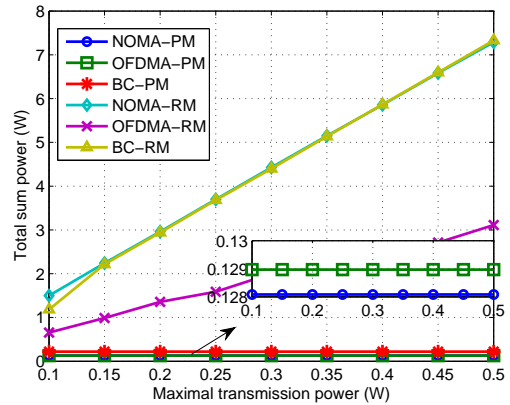}
\caption{Sum power of the system versus the maximal transmission power. \label{fig16}}
\end{figure}

We try multiple starting points in the simulations to exhaustively obtain a near globally optimal solution.
We test 20 randomly generated channels shown in Fig.~\ref{fig19}, where NOMA-RM-EXH refers to the NOMA-RM algorithm with 1000 starting points for each channel realization.
It can be seen that the sum rate of NOMA-RM is almost the same as that of NOMA-RM-EXH, which indicates that the proposed NOMA-RM approaches the near globally optimal solution.

Fig.~\ref{fig16} shows the sum power versus the maximal transmission power under various algorithms.
It is observed that the sum power of NOMA/OFDMA/BC-RM increases with the transmission power constraint.
This is because increasing the overall transmission power is always beneficial in enhancing the sum rate of the system.
It is also found that the sum power keeps the same for NOMA/OFDMA/BC-PM.
This is due to that the maximal transmission power for each BS is set as the same and the sum power does not change value with the maximal transmission power for power minimization.
From Fig.~\ref{fig16}, the NOMA-PM is better than OFDMA/BC-PM in terms of the sum power consumption.
The reason is that NOMA applies SIC to utilize intra-cell interference and each user can occupy the total available bandwidth, which results in lower sum power than OFDMA and BC.

\begin{figure}
\centering
\includegraphics[width=3.0in]{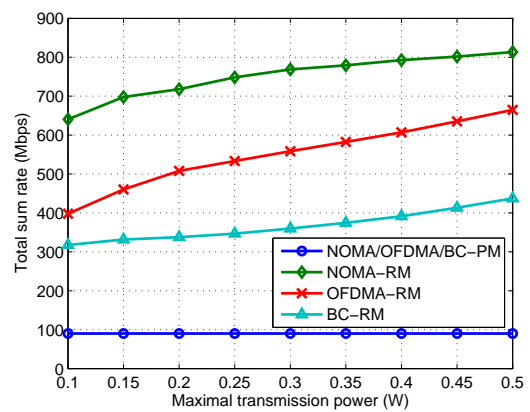}
\caption{Sum rate of the system versus the maximal transmission power.\label{fig17}}
\end{figure}

\begin{figure}
\centering
\includegraphics[width=3.0in]{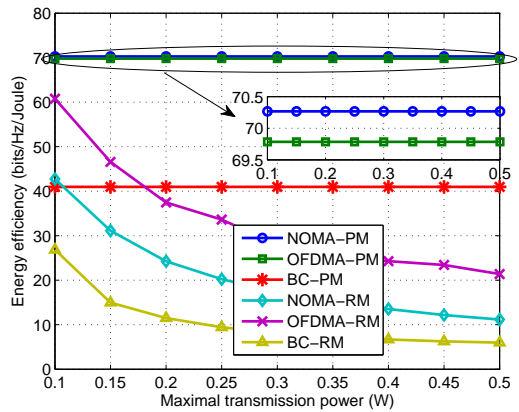}
\caption{Energy efficiency of the system versus the maximal transmission power. \label{fig18}}
\end{figure}

We illustrate the sum rate and energy efficiency (the ratio of sum rate and sum power) versus the maximal transmission power under various algorithms in Fig.~\ref{fig17} and Fig.~\ref{fig18}, respectively.
It is seen from Fig.~\ref{fig17} that the sum rate is the same for NOMA/OFDMA/BC-PM.
Due to power minimization, each user is set as satisfying the minimal rate demand.
Therefore, the sum rate of NOMA/OFDMA/BC-PM keeps a constant, i.e., $2IMR$, as shown in Fig.~\ref{fig17}.
From Fig.~\ref{fig17}, it is observed that NOMA-RM outperforms OFDMA/BC-RM in terms of sum rate.
One reason is that each user in NOMA networks can be allocated with higher fraction of bandwidth than in OFDMA networks, where the bandwidth is orthogonally distributed to different users in the same cell.
The other reason is that NOMA can efficiently  utilize intra-cell interference by using SIC, which results in higher rate than that of BC.
From Fig.~\ref{fig18}, it is interesting to observe that NOMA-PM achieves the best energy efficiency among all algorithms.
It is also found that the energy efficiency of NOMA/OFDMA/BC-RM monotonically decreases with the maximal transmission power.
This is because sum rate maximization algorithms tend to transmit with large power from Fig.~\ref{fig16}, which results in large intra/inter-interference and low energy efficiency.
NOMA needs to broadcast additional information about the decoding orders and the receivers at the users are complicated compared with OFDMA/BC-PM.
Thus, we can conclude that NOMA achieves some performance gains at the cost of some additional information broadcasting of the BS and extra computations of the users from Fig.~\ref{fig16} and Fig.~\ref{fig17}.

\section{Conclusion}

In this paper, we aim at sum power minimization and sum rate maximization through power control for multi-cell networks with NOMA.
Both sum power minimization and sum rate maximization problems can be transformed into correspondingly equivalent problems with smaller variables.
For sum power minimization, users with poor channel gains tend to be allocated with more power.
Sum rate maximization problem can be decoupled into two subproblems, i.e., power allocation problem in a single cell and power control problem in multiple cells.
The power allocation problem in a single cell is proved to be convex and its globally optimal solution can be obtained in the closed-form expression.
Based on the optimal solution to power allocation problem in a single cell, only user in each cell with the best channel gain deserves additional power from its served BS to maximize sum rate in multi-cell networks with NOMA.
Through simulation results, it tends to pair users with distinctive channel gains for both sum power minimization and sum rate maximization.
It is shown that the proposed power control methods can be applied to MIMO systems.
It is also verified that NOMA outperforms OFDMA and BC in terms of sum power minimization and sum rate maximization at the cost of some additional information broadcasting of the BSs and computations of the users.
The users' mobility
issue for multi-cell NOMA systems is left for our future work.

\appendices
\section{Proof of Theorem 1}

\setcounter{equation}{0}
\renewcommand{\theequation}{\thesection.\arabic{equation}}


To prove Theorem 1,
we find that the objective function (\ref{power_mini1_obj1}) is only a function of $\pmb q$ and variable $\pmb p$ only exists in constraints (\ref{max1}b), (\ref{max1}c) and $\pmb p \geq \pmb 0$.
Hence, $\pmb p$ can be viewed as an intermediate variable.
With this observation, sum power minimization Problem (\ref{max1}) with objective function (\ref{power_mini1_obj1}) can be simplified by removing $\pmb p$ without loss of optimality.
Specifically, the constraints (\ref{max1}b), (\ref{max1}c) and $\pmb p \geq \pmb 0$ about variable $\pmb p$ can be equivalently transformed into constraints about $\pmb q$.

Given total transmission power of other BSs, inter-cell interference is fixed, hence Problem (\ref{max1}) with objective function (\ref{power_mini1_obj1}) can be readily simplified into a single-cell power allocation problem.
For BS $i$, the power allocation problem with given $\pmb q_{-i}=[q_{11}, \cdots, q_{(i-1)M}, q_{(i+1)1}, \cdots,  q_{IM}]^T$ can be formulated as
\begin{subequations}\label{max2_2}
\begin{align}
\mathop{\min}_{\pmb{p}_i\geq \pmb 0, \pmb q_i\geq \pmb 0}\;
 \quad&  \sum_{m=1}^M q_{im}\\
\textrm{s.t.}\quad\qquad \!\!\!\!\!\!\!\!\!
&
\sum_{j=J_{i(m-1)}+1}^{J_{im}}p_{ijm}=q_{im},\quad \forall m \in \mathcal M\\
&
 p_{ijm}
 \geq \left(2^{\frac{R_{ijm}}B} -1\right)\left(
 \sum_{n=j+1}^{J_{im}} p_{inm}+
H_{ijm}
 \right),
 \nonumber \\
  &\qquad
  \quad\forall m\in\mathcal M, j\in \mathcal J_{im},
\end{align}
\end{subequations}
where $\pmb p_i=[p_{i(J_{i0}+1)1}, \cdots, p_{iJ_{iM}M}]^T$, and $\pmb q_i=[q_{i1}, \cdots, q_{iM}]^T$.

Observing that both the objective function and constraints of Problem (\ref{max2_2}) can be decoupled, Problem (\ref{max2_2}) can be further decoupled into multiple single-subchannel problems.
We consider the following optimization problem on subchannel $m$:
\begin{subequations}\label{max2_2appenA}
\begin{align}
\mathop{\min}_{\pmb{p}_{im}\geq \pmb 0}\;
 \quad&  \sum_{j=J_{i(m-1)}+1}^{J_{im}}p_{ijm}\\
\textrm{s.t.}\quad\qquad \!\!\!\!\!\!\!\!\!
&
 p_{ijm}
 \geq \left(2^{\frac{R_{ijm}}B} -1\right)\left(
 \sum_{n=j+1}^{J_{im}} p_{inm}+
H_{ijm}
 \right),
 \nonumber \\
  &\qquad
  \quad\forall j\in \mathcal J_{im},
\end{align}
\end{subequations}
where $\pmb p_{im}=[p_{i(J_{i(m-1)}+1)m}, \cdots, p_{iJ_{im}m}]^T$.
Combining (\ref{max2_2appenA}b) and $\pmb{p}_{im}\geq \pmb 0$, we find that constraints (\ref{max2_2appenA}b) hold with equality for any optimal solution to Problem (\ref{max2_2appenA}), as otherwise (\ref{max2_2appenA}a) can be further improved, contradicting that the solution is optimal.
Setting constraints (\ref{max2_2appenA}b) with equality, we obtain
\begin{eqnarray}\label{appenAeq2}
2^{\frac{R_{ijm}}B}  \sum_{n=j+1}^{J_{im}} p_{inm}+
 \left(2^{\frac{R_{ijm}}B} -1\right)H_{ijm}=   \sum_{n=j}^{J_{im}} p_{inm},
\end{eqnarray}
for all $j \in \mathcal J_{im}$.
Define
\begin{equation}\label{appenAeq2_1}
b_{ijm}=\sum_{n=j}^{J_{im}}p_{inm}, \quad j\in\mathcal  J_{im},
\end{equation}
which represents the summation of the transmission power from user $j$ to user $J_{im}$.
Based on (\ref{appenAeq2}) and (\ref{appenAeq2_1}), we can obtain
\begin{equation}\label{appenAavalue2_3}
{b_{ijm}}={2^{\frac{R_{ijm}}B}} b_{i({j+1})m}+
 { \left(2^{\frac{R_{ijm}}B} -1\right)H_{ijm}}, \quad \forall j\in \mathcal J_i .
\end{equation}
Defining $b_{i(J_{im}+1)m}=0$ and ${2^{\sum_{s=J_{im}}^{J_{im}-1}\frac{R_{ism}}B}}=2^0$, we can rewrite (\ref{appenAavalue2_3}) as follows:
\begin{equation}\label{appenAeqbvalue}
b_{ijm}=\sum_{n=j}^{J_{im}}
 { \left(2^{\frac{R_{inm}}B} -1\right)}
 {2^{\sum_{s=j}^{n-1}\frac{R_{ism}}B}}H_{inm}, \quad j\in \mathcal J_{im}.
\end{equation}
According to (\ref{appenAeq2_1}), we know $p_{ijm}=b_{ijm}-b_{i(j+1)m},  \forall j \in \mathcal J_{im}$.
Substituting (\ref{appenAeqbvalue}) into $p_{ijm}=b_{ijm}-b_{i(j+1)m}$ yields
\begin{equation}\label{appenAeq3}
\begin{aligned}
p_{ijm}& =
\sum_{n=j}^{J_{im}}
{ \left(2^{\frac{R_{inm}}B} -1\right)}
{2^{\sum_{s=j}^{n-1}\frac{R_{ism}}B}}H_{inm}
 -\\
 &
\sum_{n=j+1}^{J_{im}}
{ \left(2^{\frac{R_{inm}}B} -1\right)}
{2^{\sum_{s=j+1}^{n-1}\frac{R_{ism}}B}}H_{inm},  \quad \forall j \in \mathcal J_{im},
\end{aligned}
\end{equation}
which is the optimal solution to Problem (\ref{max2_2appenA}).

From (\ref{appenAeq2_1}) and (\ref{appenAeqbvalue}), the optimal objective value of Problem (\ref{max2_2appenA}) is
\begin{eqnarray}\label{appenAeq4}
q_{im}
&&\!\!\!\!\!\!\!\!\!\!
=
\sum_{j=J_{i(m-1)}+1}^{J_{im}}p_{ijm}
\nonumber\\
&&\!\!\!\!\!\!\!\!\!\!
=\sum_{n=J_{i(m-1)}+1}^{J_{im}}
{ \left(2^{\frac{R_{inm}}B}-1\right)}
{2^{\sum_{s=J_{i(m-1)}+1}^{n-1}\frac{R_{ism}}B}}H_{inm},\nonumber\\
\end{eqnarray}
which is minimal sum power of $\sum_{j=J_{i(m-1)}+1}^{J_{im}} p_{ijm}$ satisfying (\ref{max2_2appenA}b) and $\pmb{p}_{im}\geq \pmb 0$.
Applying (\ref{appenAeq4}) into Problem (\ref{max1}) with objective function (\ref{power_mini1_obj1}) yields equivalent Problem (\ref{max2}),
where the inequality shown in (\ref{max2}b) is due to the fact that (\ref{appenAeq4}) is the minimal value of $q_{im}$.

\section{Proof of Theorem 2}
\setcounter{equation}{0}
\renewcommand{\theequation}{\thesection.\arabic{equation}}

We prove each of the three properties required for standard function below.

Positivity: Since $\max\limits_{l\in\{j, \cdots, J_i\}}\frac{\sigma^2}{|h_{ilm}|^2}>0$ for $j \in \mathcal J_{im}$, we have $f_{im}(\pmb q)> 0$ from (\ref{max2}b).

Monotonicity: Let total transmission power vector $\pmb{q}^{(1)}=[q_{11}^{(1)}, \cdots, q_{IM}^{(1)}]^T$
and $\pmb{q}^{(2)}=[q_{11}^{(2)}, \cdots, q_{IM}^{(2)}]^T$ be such
that $q_{im}^{(1)} \geq q_{im}^{(2)}$, $\forall i \in \mathcal{I}, m \in \mathcal M$.
Then, we have
\begin{eqnarray}\label{AppenBIntFunceq1}
&&\!\!\!\!\!\!\!\!\!\!\max_{l\in\{j, \cdots, J_i\}}\frac{\sum_{k \in \mathcal I \setminus \{i\}} q_{km}^{(1)} |h_{klm}|^2+\sigma^2}{|h_{ilm}|^2}
\nonumber\\
&&\!\!\!\!\!\!\!\!\!\!\geq
\max_{l\in\{j, \cdots, J_{im}\}}\frac{\sum_{k \in \mathcal I \setminus \{i\}} q_{km}^{(2)} |h_{klm}|^2+\sigma^2}{|h_{ilm}|^2}.
\end{eqnarray}
According to (\ref{AppenBIntFunceq1}) and (\ref{max2}b), we have $f_{im}(\pmb q^{(1)})\geq f_{im}(\pmb q^{(2)})$.

Scalability: Letting $\pmb{q} \geq \pmb{0}$ and $\lambda >1$, we have
\begin{eqnarray}\label{AppenBIntFunceq2}
&&\!\!\!\!\!\!\!\!\!\!\max_{l\in\{j, \cdots, J_{im}\}}\lambda\frac{\sum_{k \in \mathcal I \setminus \{i\}} q_{km} |h_{klm}|^2+\sigma^2}{|h_{ilm}|^2}
\nonumber\\
&&\!\!\!\!\!\!\!\!\!\!>
\max_{l\in\{j, \cdots, J_{im}\}}\frac{\sum_{k \in \mathcal I \setminus \{i\}} \lambda q_{km} |h_{klm}|^2+\sigma^2}{|h_{ilm}|^2},
\end{eqnarray}
where the inequality follows from that $\sigma^2>0$.
Based on (\ref{AppenBIntFunceq2}) and (\ref{max2}b), we have $\lambda f_{im}(\pmb q)> f_{im}(\lambda \pmb q)$.

\section{Proof of Theorem 3}
\setcounter{equation}{0}
\renewcommand{\theequation}{\thesection.\arabic{equation}}

Since the constraints of  Problem (\ref{ratemax3}) are all linear, we only need to prove that the objective function  (\ref{ratemax3}a) is convex.
We first rewrite (\ref{ratemax3}a) as
\begin{eqnarray}
R_{im}=
&&\!\!\!\!\!\!\!\!\!
-B\log_2\left(
{ \sum_{n=J_{i(m-1)}+1}^{J_{im}} p_{inm}
+
H_{i(J_{i(m-1)}+1)m}}
 \right)
 \nonumber\\
&&\!\!\!\!\!\!\!\!\!
 +
\sum_{j=J_{i(m-1)}+2}^{J_{im}}
B \left(\log_2\left(
{ \sum_{n=j}^{J_{im}} p_{inm}  +
H_{i(j-1)m}} \right)
\right.
\nonumber\\
&&\!\!\!\!\!\!\!\!\!\!\left.
-
\log_2\left(
{ \sum_{n=j}^{J_{im}} p_{inm} +
H_{ijm}}  \right) \right) +
B\log_2(H_{iJ_{im}m})
,\nonumber
\end{eqnarray}
where $R_{im}$ is negative sum rate of all users served by BS $i$ on subchannel $m$.
The second-order derivative of $R_{im}$ equals
\begin{eqnarray}
&&\!\!\!\!\!\!\!\!\!\frac{\partial^2 R_{im}} {\partial p_{ijm}^2}
=
 \sum_{l= J_{i (m- 1)}+2}^{j}
\frac{1}{\ln2}\left(
\frac{B }
{\left(
{  \sum_{n =  l}^{J_{im}} p_{inm} +
H_{ijm}} \right)^2 }
\right.
\nonumber\\
&&\!\!\!\!\!\!\!\!\!\!\!\left.
-
\frac{B }
{
\left(
{  \sum_{n=l}^{J_{im}} p_{inm} +
H_{i(j-1)m}}  \right)^2}
 \right)
\nonumber\\
&&\!\!\!\!\!\!\!\!\!\!\!
+
 \frac{B }
{(\ln2)\left(  \sum_{n=J_{i(m-1)}+1}^{J_{im}} p_{inm}+ H_{i(J_{i(m-1)}+1)m}
\right)^2}
\label{secondorderder1}
\end{eqnarray}
for all $ j=J_{i-1}+1, J_{i-1}+2, \cdots, J_i$,
and
\begin{eqnarray}
&&\!\!\!\!\!\!\!\!\!\frac{\partial^2 R} {\partial p_{ijm} \partial p_{ilm}}
=
 \sum_{l= J_{i (m- 1)}+2}^{j}
\frac{1}{\ln2}\left(
\frac{B }
{\left(
{  \sum_{n =  l}^{J_{im}} p_{inm} +
H_{ijm}} \right)^2 }
\right.
\nonumber\\
&&\!\!\!\!\!\!\!\!\!\!\!\left.
-
\frac{B }
{
\left(
{  \sum_{n=l}^{J_{im}} p_{inm} +
H_{i(j-1)m}}  \right)^2}
 \right)
\nonumber\\
&&\!\!\!\!\!\!\!\!\!\!\!
+
 \frac{B }
{(\ln2)\left(  \sum_{n=J_{i(m-1)}+1}^{J_{im}} p_{inm}+ H_{i(J_{i(m-1)}+1)m}
\right)^2}
\label{secondorderder2}
\end{eqnarray}
for all $J_{i-1}+\leq j <l \leq J_i$.
Comparing (\ref{secondorderder1}) and (\ref{secondorderder2}), we find that $\frac{\partial^2 R_{im}} {\partial p_{ijm}\partial p_{ilm}}=\frac{\partial^2 R_{im}} {\partial p_{ijm}^2}$ for any $J_{i(m-1)}+1\leq j\leq l\leq J_{im}$.
Therefore, denoting $a_{ijm}=\frac{\partial^2 R_{im}} {\partial p_{ijm}^2}$ for notational simplicity, the Hessian matrix
$\pmb A_{im}$ of (\ref{ratemax3}a) has the following structure:
\begin{equation}\label{secondHess}
 \pmb A_{im} \!\!=\!
\!\!\begin{pmatrix}
\!a_{i(J_{i(m-1)}+1)m}\! & \!\!\!a_{i(J_{i(m-1)}+1)m} \! \!\! & \!\cdots \!& \!\!\! a_{i(J_{i(m-1)}+1)m}\! \\
\!a_{i(J_{i(m-1)}+1)m} \!& \!\!\!a_{i(J_{i(m-1)}+2)m} \! \!\! & \!\cdots \!& \!\!\!a_{i(J_{i(m-1)}+2)m}\! \\
 \vdots & \vdots  \ &  & \vdots\\
\!a_{i(J_{i(m-1)}+1)m}\! & \!\!\!a_{i(J_{i(m-1)}+2)m} \!\!\!  & \!\cdots\! & \!\!\!a_{iJ_{im}m}\!
 \end{pmatrix}\!\! .
\end{equation}

%
Based on (\ref{secondHess}),
the $t$-th order principal minor of matrix $\pmb A_{im}$ can be expressed as
\begin{eqnarray}
A_{imt}\!\!&&\!\!\!\!\!\!\!\!\! =\!
\begin{vmatrix}
\!a_{i(J_{i(m-1)}+1)m}\!\! & \!\!a_{i(J_{i(m-1)}+1)m}\!\! & \!\cdots\! & \!\!a_{i(J_{i(m-1)}+1)m} \!\\
\!a_{i(J_{i(m-1)}+1)m}\!\! & \!\!a_{i(J_{i(m-1)}+2)m}\!\! & \!\cdots \!& \!\!a_{i(J_{i(m-1)}+2)m} \!\\
\vdots & \vdots   &  & \vdots\\
\!a_{i(J_{i(m-1)}+1)m}\!\! & \!\!a_{i(J_{i(m-1)}+2)m}\!\! &\! \cdots\! &\!\! a_{i(J_{i(m-1)}+t)m} \!\\
\end{vmatrix}\nonumber
\\
&&\!\!\!\!\!\!\!\!\!\!\!
=a_{i(J_{i(m-1)}+1)m} \prod_{l=2}^t (a_{i(J_{i(m-1)}+l)m} -a_{i(J_{i(m-1)}+l-1)m} ).\nonumber\\
\label{hessianT}
\end{eqnarray}
Since
\begin{eqnarray*}
a_{i(J_{i(m-1)}+1)m} &&\!\!\!\!\!\!\!\!\!\!=\!
\frac{B /(\ln2)}
{\left( \sum_{n=J_{i(m-1)}+1}^{J_{im}} p_{inm}+
H_{i(J_{i(m-1)}+1)m}
\right)^2}
\\&&\!\!\!\!\!\!\!\!\!\!
>0,
\end{eqnarray*}
and for $J_{i(m-1)}+2 \leq l \leq J_{im}$,
\begin{eqnarray*}
 &&\!\!\!\!\!\!\!\!\!\!\!\!\!\!\!\!\! \quad
a_{i(J_{i(m-1)}+l)m} -a_{i(J_{i(m-1)}+l-1)m}
\nonumber\\
 &&\!\!\!\!\!\!\!\!\!\!\!\!\!\!\!\!\! \quad
=
\frac{1}{\ln2} \left(
\frac{B }
{\left(
{  \sum_{n = J_{i(m-1)}+l}^{J_{im}} p_{inm} +
H_{i(J_{i(m-1)}+l)m}}  \right) ^2 }
\right.
\nonumber\\
 &&\!\!\!\!\!\!\!\!\!\!\!\!\!\!\!\!\! \qquad
\left.
-
\frac{B }
{
\left(
{  \sum_{n=J_{i(m-1)}+l}^{J_{im}} p_{inm}  +
H_{i(J_{i(m-1)}+l-1)m}} \right) ^2}
\right)
\overset{(\text {a})}{\geq} 0,
\end{eqnarray*}
where inequality (a) holds based on (\ref{power_min2_22}),
we have from (\ref{hessianT}) that $A_{imt} \geq 0$ for $t=1, \cdots, J_{im}-J_{i(m-1)}$.
According to \cite[Page~558]{horn2012matrix}, a function whose Hessian is positive semi-definite throughout a convex set is convex.
Besides, if the principal minors of a matrix are all nonnegative,  this matrix is positive semi-definite \cite[Page~558]{horn2012matrix}.
Thus, matrix $\pmb A_{im}$ is positive semi-definite, which implies that objective function (\ref{ratemax3}a) is convex.
As a result, Problem (\ref{ratemax3}) is convex.

Then, we prove the feasibility condition for Problem (\ref{ratemax3}).
To prove this, we denote
\begin{equation}\label{max1_8}
\tilde q_{im}= \left\{ \begin{array}{ll}
\!\! \mathop{\min}\limits_{\pmb{p}_{im}\geq \pmb 0}
&\!
\sum_{j=J_{i(m-1)}+1}^{J_{im}}p_{ijm}
\\
 \:\textrm{s.t.}
&  \!p_{ijm}
  \!\geq \! \!\left(\! 2^{\frac{R_{ijm}}B}  \!-\!1\! \right)\!\left(\!
 \sum_{n=j+1}^{J_{im}}\! p_{inm} \!+\!
H_{ijm}\!
  \right),
 \nonumber\\
 &
 \qquad\forall j\in \mathcal J_{im}.
\end{array} \right.\nonumber
\end{equation}
From (\ref{ratemax3}b), (\ref{ratemax3}c), and $\pmb p_{im}\geq \pmb 0$, we can find that Problem (\ref{ratemax3}) is feasible if and only if $\tilde q_{im}\leq q_{im}$.
To obtain $\tilde  q_{im}$, we observe that constraints (\ref{ratemax3}c) hold with equality for all $j\in\mathcal J_{im}$, as otherwise $\tilde q_{im}$ can be further improved.
By solving these $J_{im}-J_{i(m-1)}$ linear equations, we have
$\tilde q_{im}=\sum_{n=J_{i(m-1)}+1}^{J_{im}}
{ \left(2^{\frac{R_{inm}}B} -1\right)}
{2^{\sum_{s=J_{i(m-1)}+1}^{n-1}\frac{R_{ism}}B}}H_{inm}
$
from (\ref{appenAeq2}), (\ref{appenAeq2_1}) and (\ref{appenAeq4}).
Hence, the feasibility condition for Problem (\ref{ratemax3}) is achieved as (\ref{ratemaxfeasiblecond}).

\section{Proof of Theorem 4}
\setcounter{equation}{0}
\renewcommand{\theequation}{\thesection.\arabic{equation}}

The Lagrangian function of Problem (\ref{ratemax3}) can be written by
\begin{eqnarray*}
&&\!\!\!\!\!\!\!\!
\mathcal L(\pmb{p}_{im}, \alpha_{im}, \pmb \beta_{im}, \pmb \gamma_{im})
 =
\alpha_{im}\!\left(\!\sum_{j=J_{i(m-1)}+1}^{J_{im}}p_{ijm}-q_{im}
 \right)
\nonumber\\
&&\!\!\!\!\!\!\!\!
  - \sum_{j=J_{i(m-1)}+1}^{J_{im}}
 B\log_2 \left(
1+\frac{  p_{ijm}}
{  \sum_{n=j+1}^{J_{im}} p_{inm}+
H_{ijm}}
 \right)
\nonumber\\
&&\!\!\!\!\!\!\!\!
+\sum_{j=J_{i(m-1)}+1}^{J_{im}} \beta_{ijm}
\left(
 \left(2^{\frac{R_{ijm}}B} -1\right)\left(
 \sum_{n=j+1}^{J_{im}} p_{inm}+
H_{ijm}
 \right) \right.
\nonumber\\
&&\!\!\!\!\!\!\!\!
-  p_{ijm}\Bigg)
-\sum_{j=J_{i(m-1)}+1}^{J_{im}} \gamma_{ijm}p_{ijm},
\end{eqnarray*}
where
$\alpha_{im}$, $\pmb \beta_{im} =[\beta_{i(J_{i(m-1)}+1)m},   \cdots, \beta_{iJ_{im}m}]^T\geq \pmb 0$ and $\pmb \gamma_{im}=[\gamma_{i(J_{i(m-1)}+1)m},  \cdots, \gamma_{iJ_{im}m}]^T\geq \pmb 0$ are the  Lagrange multipliers associated with the corresponding constraints of Problem (\ref{ratemax3}).
The KKT conditions of Problem (\ref{ratemax3}) are:
\begin{subequations}\label{appenBkktcond}
\begin{align}
\!\!\!\!\!\!\!\!\!\!\!\!\!\!\!\!&
\frac{\partial \mathcal L}{\partial p_{ijm}}=
-\frac{B }
{(\ln2)\left(  \sum_{n=J_{i(m-1)}+1}^{J_{im}} p_{inm}+H_{i(J_{i(m-1)}+1)m}
\right)}
\nonumber\\
&\qquad+
\alpha_{im}+\sum_{n=J_{i(m-1)}+1}^{j-1}\left(2^{\frac{R_{inm}}{B}}-1\right)\beta_{inm}
-\beta_{ijm}\nonumber\\
&
\qquad -\gamma_{ijm}- \sum_{l= J_{i(m - 1)}+ 2}^{j}
\frac{1}{\ln2}\left(
\frac{B }
{
{  \sum_{n =  l}^{J_{im}} p_{inm}+
H_{ijm}} }
\right.
\nonumber\\
&\qquad
\left.
-
\frac{B }
{
{  \sum_{n=l}^{J_{im}} p_{inm} +
H_{i(j-1)m}}}
\right), 
\quad \forall j \in \mathcal J_{im}
\\
 &
\beta_{ijm}
\left(
 \left(2^{\frac{R_{ijm}}B} -1\right)\left(
 \sum_{n=j+1}^{J_{im}} p_{inm}+
H_{ijm}
 \right)
 -  p_{ijm}
 \right)=0,
 \nonumber\\
 &\qquad
 \quad\forall j\in \mathcal J_{im} \\
 &\gamma_{ijm} p_{ijm}=0, \quad \forall j \in \mathcal J_{im}\\
 &\sum_{j=J_{i(m-1)}+1}^{J_{im}}p_{ijm}-q_{im}=0\\
 &
 \left( 2^{\frac{R_{ijm}}B}  - 1 \right)\left(
  \sum_{n=j+1}^{J_{im}} p_{inm} +
H_{ijm}
  \right) -  p_{ijm} \leq 0,
  \nonumber\\
& \qquad \forall j\in \mathcal J_{im}\\
& \pmb \beta_{im}, \pmb \gamma_{im}, \pmb p_{im} \geq \pmb 0.
\end{align}
\end{subequations}
According to (\ref{appenBkktcond}e) and $R_{ijm}>0$, we can obtain $p_{ijm}>0$, $\forall j \in \mathcal J_{im}$.
Hence, further combing (\ref{appenBkktcond}c), we have
\begin{equation}\label{appenBeq1}
\gamma_{ijm}=0, \quad \forall j \in \mathcal J_{im}.
\end{equation}

Assume that
\begin{equation}\label{appenBsys2_1}
 {H_{i(J_{i(m-1)}+1)m}}
> {H_{i(J_{i(m-1)}+2)m}}
>\cdots>{H_{i J_{im}m}}.
\end{equation}
The special case with $ {H_{i(j-1)m}}={H_{ijm}}$ for at least one $j\in \{J_{i(m-1)}+2, \cdots, J_{im}\}$ is considered later.
From (\ref{appenBkktcond}a) and (\ref{appenBeq1}), we obtain
\begin{eqnarray*}
\frac{\partial \mathcal L}{\partial p_{ijm}}-\frac{\partial \mathcal L}{\partial p_{i(j-1)m}}
&&\!\!\!\!\!\!\!\!\!=
\frac{1}{\ln2}\left(
\frac{B }
{
{  \sum_{n=j}^{J_{im}} p_{inm} + H_{i(j-1)m}}}
\right.\nonumber\\
&&\!\!\!\!\!\!\!\!\!\!\!
\left.
-
\frac{B }
{
{  \sum_{n =  j}^{J_{im}} p_{inm} +H_{ijm}} }
\right)
 \nonumber\\
&&\!\!\!\!\!\!\!\!\!\!\!
 -\beta_{ijm}
+ \beta_{i(j-1)m} 2^{\frac{R_{i(j-1)m}}{B}}
=0\label{appenBkkt2}
\end{eqnarray*}
for $j=J_{i(m-1)}+2, \cdots, J_{im}$.
Considering (\ref{appenBsys2_1}), we have
\begin{eqnarray*}
&&\!\!\!\!\!\!\!\!\!
\beta_{ijm} - \beta_{i(j-1)m} 2^{\frac{R_{i(j-1)m}}{B}}
\nonumber\\
&&\!\!\!\!\!\!\!\!\!
\!=\!\frac{1}{\ln2}\!\left(\!
\frac{B}
{
{ \sum_{n=j}^{J_{im}} p_{inm} \! +\!
 {H_{i(j-1)m}}
}}
\! -\!
\frac{B}
{\!
{ \sum_{n =  j}^{J_{im}} p_{inm} \!+\!
 {H_{ijm}}
  }}
 \right)
<0
\label{appenBkkt3}
\end{eqnarray*}
for $j=J_{i(m-1)}+2, \cdots, J_{im}$.
Since $\beta_{ijm}\geq 0$, we have $\beta_{i(j-1)m} 2^{\frac{R_{i(j-1)m}}{B}} >\beta_{ijm} \geq 0$, $j=J_{i(m-1)}+2, \cdots, J_{im}$.
Thus, we can obtain $\beta_{i(J_{im}-1)m}>0$, $\cdots$, $\beta_{i(J_{i(m-1)}+1)m}>0$.
Hence,
we only need to consider the following two cases of $\beta_{i J_{im}m}$ for user $J_{im}$.

1) If $\beta_{iJ_{im}m}>0$,
      constraints in (\ref{appenBkktcond}b) are satisfied via $\left(2^{\frac{R_{ijm}}B} -1\right)\left(\sum_{n=j+1}^{J_{im}} p_{inm}+H_{ijm}
 \right)
 -  p_{ijm}=0$ for all $j\in\mathcal  J_{im}$, which implies that the minimal rate constraints (\ref{appenBkktcond}e) hold with equality for all users.
Thus, the optimal value of Problem (\ref{ratemax3}) is $-\sum_{j=J_{i(m-1)}+1}^{J_{im}} R_{ijm}$, and the optimal solution to Problem (\ref{ratemax3}) can be obtained as in (\ref{power_mini2}) by solving constraints (\ref{appenBkktcond}e) with equality for all users.

2) If $\beta_{iJ_{im}m}=0$, we find that constraints (\ref{appenBkktcond}e) hold with equality except for the user $J_{im}$.

Due to that Problem (\ref{ratemax3}) can be easily solved for the case $\beta_{iJ_{im}m}>0$, we only need to consider the case $\beta_{iJ_{im}m}=0$ in the following.
Since minimal rate constraints (\ref{appenBkktcond}e) hold with equality for $j\in\mathcal J_{im}\setminus\{J_{im}\}$, we find that the additional power is allocated to the user with the highest channel gain and other users served by BS $i$ on subchannel $m$ are allocated with minimal transmission power to meet the minimal rate demands.
Now, it remains optimal to solve constraints (\ref{appenBkktcond}e) with equality for $j\in\mathcal J_{im}\setminus\{J_{im}\}$.
Thus, we have
\begin{eqnarray}\label{appenBavalue2}
2^{\frac{R_{ijm}}B}  \sum_{n=j+1}^{J_{im}} p_{inm}+
 \left(2^{\frac{R_{ijm}}B} -1\right)
H_{ijm}=   \sum_{n=j}^{J_{im}} p_{inm},
\end{eqnarray}
for all $j=J_{i(m-1)}+1, \cdots, J_{im}-1$.
Define
\begin{equation}\label{appenBavalue2_1}
b_{ijm}=\sum_{n=j}^{J_{im}}p_{ijm}, \quad j=J_{i(m-1)}+1, \cdots, J_{im}.
\end{equation}
Substituting (\ref{appenBavalue2_1}) into (\ref{appenBkktcond}d) yields
\begin{equation}\label{appenBavalue2_2}
b_{i(J_{i(m-1)}+1)}=q_{im}.
\end{equation}
Based on (\ref{appenBavalue2}) and (\ref{appenBavalue2_1}), we can obtain
\begin{equation}\label{appenBavalue2_3}
b_{i({j+1})m}=\frac{b_{ijm}}{2^{\frac{R_{ijm}}B}}-
\frac{ \left(2^{\frac{R_{ijm}}B} -1\right)H_{ij}}
{2^{\frac{R_{ijm}}B}  },
\end{equation}
for all $j=J_{i(m-1)}+1, \cdots, J_{im}-1$.
By further using (\ref{appenBavalue2_2}), we have
\begin{eqnarray}\label{avalue}
b_{i(j+1)m} =&&\!\!\!\!\!\!\!\! \frac{q_{im}}{ 2^{\sum_{n=J_{i(m-1)}+1}^{j}\frac{R_{inm}}B}}
\nonumber\\
&&\!\!\!\!\!\!\!\!- \sum_{n=J_{i(m-1)}+1}^{j}\frac{\left(2^{\frac{R_{inm}}B} - 1\right)
H_{inm}}{ 2^{\sum_{l=n}^{j}\frac{R_{ilm}}B}},
\end{eqnarray}
for all $j=J_{i(m-1)}+1, \cdots, J_{im}-1$.
From (\ref{appenBavalue2_1}), we can obtain
\begin{equation}\label{powervalue}
\begin{aligned}
& p_{ijm}\!=\!
  \begin{cases}
 b_{ijm}-b_{i(j+1)m}, & \!\! \!\!\!\mbox{ if $j=J_{i(m-1)}+1, \cdots, J_{im}-1$}\\
b_{ijm}, & \!\!\!\!\!\mbox{ if $j=J_{im}$}
  \end{cases}.
\end{aligned}
\end{equation}
By inserting (\ref{avalue})  into (\ref{powervalue}), we can obtain closed-form expression of $p_{ijm}$ as (\ref{ratemaxeq1}).
Substituting (\ref{ratemaxeq1}) into objective function (\ref{ratemax3}a), we can obtain the optimal sum rate of Problem (\ref{ratemax3})  as (\ref{ratemaxeq2}).

Now, we consider the special case remained to be discussed.
Assume that there are two users served by BS $i$ with satisfying $  {H_{i(j-1)m}}
=  {H_{ijm}}$.
In this case, we can define a new user $j'$ with $R_{ij'm}=R_{i(j-1)m}+R_{ijm}$.
Calculate the optimal power allocation strategy
$[p_{i(J_{i(m-1)}+1)m}^*, \cdots, $ $p_{i(j-2)m}^*,p_{ij'm}^*, p_{i(j+1)m}^*, \cdots, p_{iJ_{im}m}^*]^T$ for users $J_{i(m-1)}+1, \cdots, j-2, j', j+1,\cdots, J_{im}$ according to (\ref{ratemaxeq1}).
Based on (\ref{sys3}),
we have
\begin{eqnarray*}
r_{i(j-1)m}+r_{ijm}&&\!\!\!\!\!\!\!\!\!\! =
B\log_2\left(
 \frac{  \sum_{n=j-1}^{J_{im}} p_{inm}\!+\!
{
H_{i(j-1)m}}
}
{  \sum_{n=j}^{J_{im}} p_{inm}\!+\!
{
H_{i(j-1)m}}
}
\right)
\nonumber\\
&&\!\!\!\!\!\!\!\!\!\!
\quad
+B\log_2\left(
 \frac{  \sum_{n=j}^{J_{im}} p_{inm}\!+\!
{
H_{ijm}}
}
{  \sum_{n=j+1}^{J_{im}} p_{inm}\!+\!
{
H_{ijm}}
}
\right)
\nonumber\\
&&\!\!\!\!\!\!\!\!\!\!
=B\log_2\left(
 \frac{  \sum_{n=j-1}^{J_{im}} p_{inm}+
{
H_{ijm}}
}
{  \sum_{n=j+1}^{J_{im}} p_{inm}+
{
H_{ijm}}
}
\right),
\end{eqnarray*}
which means that the sum rate of user $j-1$ and user $j$ is determined by the sum power $p_{i(j-1)m}+p_{ijm}$.
In the optimal power allocation strategy for user $j-1$ and user $j$, we can arbitrarily allocate power $p_{i(j-1)m}^*$ and $p_{ijm}^*$ with $p_{i(j-1)m}^*+p_{ijm}^*=p_{ij'm}^*$ fixed and the minimal rate constraints $r_{i(j-1)m}^*\geq R_{i(j-1)m}$ and $r_{ijm}^*\geq R_{ijm}$ satisfied.
If $j\neq J_{im}$, we can observe that $r_{ij'm}^*=R_{ij'm}=R_{i(j-1)m}+R_{ijm}$ according to (\ref{appenBavalue2}) for the optimal power allocation strategy.
Then, $r_{i(j-1)m}^*= R_{i(j-1)m}$ and $r_{ijm}^*= R_{ijm}$, which indicates that the optimal power for user $j-1$ and user $j$ can be presented as (\ref{ratemaxeq1}).
If $j=J_{im}$, we can observe that $r_{ij'm}^*\geq R_{ij'm}$ according to (\ref{appenBkktcond}e) for the optimal power allocation strategy.
If we set $r_{i(j-1)m}^*= R_{i(j-1)m}$ and $r_{ijm}^*=r_{ij'm}^*-R_{ijm}$, the optimal power for user $j-1$ and user $j$ can also be presented as (\ref{ratemaxeq1}).
Thus, we can still obtain the optimal sum rate of Problem (\ref{ratemax3})  as (\ref{ratemaxeq2}).

As a result, Theorem 4 is proved.

\bibliographystyle{IEEEtran}
\bibliography{IEEEabrv,MMM}

\end{document}